\documentclass[aps,prl,reprint,superscriptaddress]{revtex4-2}

\usepackage{textcomp}
\usepackage{array}

\usepackage{graphicx}
\usepackage{float}
\usepackage{amsmath}
\usepackage{amssymb}
\usepackage{epstopdf}
\usepackage{bm}
\usepackage{mathtools}
\usepackage[colorlinks=true, citecolor={black}, urlcolor={black}, linkcolor = {black}]{hyperref}
\usepackage{siunitx}
\usepackage{comment}
\usepackage{bold-extra}

\begin{document}

\title{Controlling Andreev bound states with the magnetic vector potential}

\author{Christian~M.~Moehle}\altaffiliation{These authors contributed equally to this work.}
\affiliation{QuTech and Kavli Institute of Nanoscience, Delft University of Technology, 2600 GA Delft, The Netherlands}

\author{Prasanna~K.~Rout}\altaffiliation{These authors contributed equally to this work.}
\affiliation{QuTech and Kavli Institute of Nanoscience, Delft University of Technology, 2600 GA Delft, The Netherlands}

\author{Nayan~A.~Jainandunsing}
\affiliation{QuTech and Kavli Institute of Nanoscience, Delft University of Technology, 2600 GA Delft, The Netherlands}

\author{Dibyendu~Kuiri}
\affiliation{AGH University of Science and Technology, Academic Centre for Materials and Nanotechnology, 30-059 Krakow, Poland}

\author{Chung~Ting~Ke}\altaffiliation{Present Address: Institute of Physics, Academia Sinica, Taipei, 11529, Taiwan}
\affiliation{QuTech and Kavli Institute of Nanoscience, Delft University of Technology, 2600 GA Delft, The Netherlands}

\author{Di~Xiao}
\affiliation{Department of Physics and Astronomy, Purdue University, West Lafayette, Indiana 47907, USA}

\author{Candice~Thomas}
\affiliation{Department of Physics and Astronomy, Purdue University, West Lafayette, Indiana 47907, USA}

\author{Michael~J.~Manfra}
\affiliation{Department of Physics and Astronomy, Purdue University, West Lafayette, Indiana 47907, USA}
\affiliation{Elmore School of Electrical and Computer Engineering, Purdue University, West Lafayette, Indiana 47907, USA}
\affiliation{School of Materials Engineering, Purdue University, West Lafayette, Indiana 47907, USA}
\affiliation{Microsoft Quantum Lab West Lafayette, West Lafayette, Indiana 47907, USA}

\author{Michał~P.~Nowak}
\affiliation{AGH University of Science and Technology, Academic Centre for Materials and Nanotechnology, 30-059 Krakow, Poland}

\author{Srijit~Goswami}
\affiliation{QuTech and Kavli Institute of Nanoscience, Delft University of Technology, 2600 GA Delft, The Netherlands}
\email{s.goswami@tudelft.nl}

\begin{abstract}	
Tunneling spectroscopy measurements are often used to probe the energy spectrum of Andreev bound states (ABSs) in semiconductor-superconductor hybrids. Recently, this spectroscopy technique has been incorporated into planar Josephson junctions (JJs) formed in two-dimensional electron gases, a potential platform to engineer phase-controlled topological superconductivity. Here, we perform ABS spectroscopy at the two ends of planar JJs and study the effects of the magnetic vector potential on the ABS spectrum. We show that the local superconducting phase difference arising from the vector potential is equal in magnitude and opposite in sign at the two ends, in agreement with a model that assumes localized ABSs near the tunnel barriers. Complemented with microscopic simulations, our experiments demonstrate that the local phase difference can be used to estimate the relative position of localized ABSs separated by a few hundred nanometers.\\
\end{abstract}

\maketitle

Hybrid structures composed of superconductors and normal conductors host Andreev bound states (ABSs)~\cite{kulik_1969,Beenakker_1991,wendin_1996_josephson}. These states are superpositions of electron-like and hole-like excitations, with energies lower than the superconducting gap. In recent years, superconductor-semiconductor hybrids have emerged as an appealing platform to manipulate these bound states. For example, controllable coupling between individual ABSs has led to the creation of Andreev molecules~\cite{su_2017, pillet_2019, kurtossy_2021, Junger_2021}, and Josephson junctions (JJs) based on these hybrids have been combined with superconducting circuits to realize Andreev qubits~\cite{Hays_2018, Hays_2021}. 

In JJs, the microscopic properties of ABSs determine global properties of the junction, such as its critical current~\cite{Beenakker_1991}. The energy of ABSs is dependent on the phase difference between the superconducting leads, which can be tuned by the application of a magnetic flux through a superconducting loop connecting the leads. In planar JJs, the vector potential of the magnetic field leads to streams of positive and negative current, to the formation of Josephson vortices, and to the well-known Fraunhofer interference pattern in critical current~\cite{Tinkham_1996, Cuevas_2007_Magnetic, Kaperek_2022}. It has been proposed that such planar JJs can host Majorana bound states~\cite{Hell_2017, Pientka_2017, Ren_2019, Fornieri_2019}, and that the location and coupling of these states can be controlled via the vector potential~\cite{Stern_2019}.

In order to investigate how the vector potential modifies ABSs in a JJ, one needs experimental techniques that provide information about the spatial extent and location of ABSs. Studies in junctions that simultaneously probe the spatial distribution and energy spectrum of ABSs have mainly been performed using scanning probe techniques~\cite{leSueur_2008, Roditchev_2015}, and more recently, via local tunnel probes in two-dimensional electron gases (2DEGs)~\cite{Banerjee_2022_topological_phase,Banerjee_2022_Local_Nonlocal}.

Here, we perform tunneling spectroscopy at both ends of planar JJs embedded in a superconducting loop, allowing us to probe the effects of the magnetic vector potential on the phase-dependence of the ABS energy. We directly show that the local superconducting phase difference originating from the vector potential has equal magnitude but opposite sign at the two ends of the JJ. This is manifested by a striking difference in the spectroscopy maps obtained from each side, in excellent agreement with a model that assumes tunnel coupling to a single ABS localized at each end. Microscopic numerical simulations confirm that such a localization of the ABSs is indeed expected, and that the tunneling current is only sensitive to ABSs located near the ends of the JJ. By modifying the potential landscape in the vicinity of the tunnel probe, we show that the local phase difference allows us to resolve multiple ABSs within a spatial extent of a few hundred nanometers, in qualitative agreement with simulations. 

\begin{figure*}[!t]
	\centering
	\includegraphics[width=1.0\textwidth]{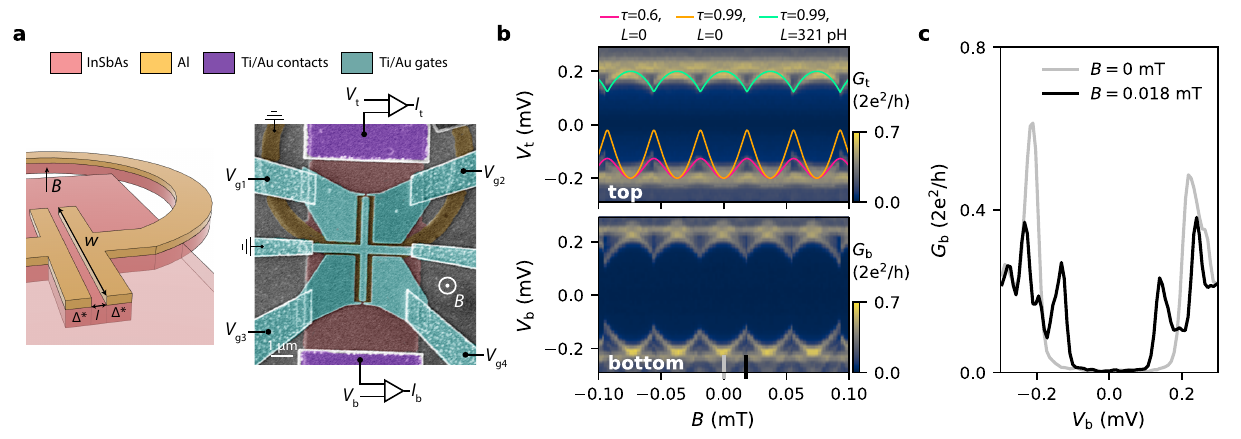}\\
	\caption{Tunneling spectroscopy at the two ends of a planar phase-biased JJ. \textbf{a} Schematic (before the gate deposition) and false-colored SEM of Dev~1. \textbf{b} Spectroscopy maps measured at the top ($V_{\mathrm{g1}}=\SI{-0.39}{V}$, $V_{\mathrm{g2}}=\SI{-0.74}{V}$, $V_{\mathrm{g3}}=\SI{0}{V}$, $V_{\mathrm{g4}}=\SI{0}{V}$) and bottom end of the JJ ($V_{\mathrm{g1}}=\SI{0}{V}$, $V_{\mathrm{g2}}=\SI{0}{V}$, $V_{\mathrm{g3}}=\SI{-1.1}{V}$, $V_{\mathrm{g4}}=\SI{-0.6}{V}$). The three curves in the top panel correspond to single-channel ABS spectra calculated for different combinations of transmission ($\tau$) and loop inductance ($L$) as specified in the legend. \textbf{c} Line cuts of the bottom spectroscopy map at the indicated positions in \textbf{b}.}
	\label{fig1}
\end{figure*}

The JJs are fabricated using an $\mathrm{InSb}_{0.92}\mathrm{As}_{0.08}$ 2DEG with in-situ grown Al as the superconductor (details about the molecular beam epitaxy growth of the heterostructure can be found in ref.~\onlinecite{Moehle_2021}). Figure \ref{fig1}a shows a schematic and a false-colored scanning electron micrograph (SEM) of such a device. To fabricate the devices, we first use a combination of Al and 2DEG etches to define the JJ and the superconducting loop. The exposed 2DEG on the top and bottom sides of the JJ is contacted by Ti/Au, and the Al loop is contacted by NbTiN, resulting in a three-terminal device. A globally deposited layer of AlO$_x$ forms the gate dielectric. Lastly, split gates are evaporated on the top and bottom ends of the JJ, allowing us to define tunnel barriers, while also depleting the 2DEG around the junction. A central gate (kept grounded throughout this study) covers the normal section of the JJ. We study two JJs (Dev~1 and Dev~2), both with length $l=\SI{80}{nm}$ and width $w=\SI{5}{\mu m}$. More details about the device fabrication can be found in the Supplementary Information - Section 1 (SI-1). The devices are measured in a dilution refrigerator with a base temperature of 30~mK using standard lock-in techniques.

In Fig.~\ref{fig1}b (top panel) we present a tunneling spectroscopy map for Dev~1 at the top end of the JJ. The conductance, $G_{\mathrm{t}}=\mathrm{d}I_{\mathrm{t}}/\mathrm{d}V_{\mathrm{t}}$, is measured as a function of voltage bias, $V_t$, and perpendicular magnetic field, $B$. The bottom panel shows the conductance measured at the bottom end, $G_{\mathrm{b}}=\mathrm{d}I_{\mathrm{b}}/\mathrm{d}V_{\mathrm{b}}$, with representative line cuts presented in Fig.~\ref{fig1}c. In both maps we see a superconducting gap that is modulated by $B$, with an oscillation period equal to $\Phi_0/S$, where $\Phi_0=h/2e$ is the magnetic flux quantum and $S$ is the area of the superconducting loop. This modulation indicates the presence of flux-periodic ABSs in the JJ. For a normal region much shorter than the superconducting coherence length, the relation between the ABS energy and the gauge-invariant phase difference between the two superconducting leads, $\varphi$, is given by~\cite{Beenakker_1991}: 

\begin{equation}
	E_n(\varphi) = \pm \Delta^{*} \sqrt{1-\tau_n \mathrm{sin}^2(\varphi/2)}, \label{eq:ABS_energy}
\end{equation}

\noindent where $\Delta^{*}$ is the induced gap in the 2DEG regions below the Al leads and $\tau_n$ is the transmission probability of the $n^\textrm{th}$ conduction channel. The flux through the loop, $\Phi=BS$, and $\varphi$ are related via $\varphi = 2\pi\Phi / \Phi_0$. The relatively small modulation depth observed in the experiment might suggest low-transmission ABSs [see the field evolution of a single ABS with $\tau=0.6$ (pink) and $\tau=0.99$ (orange) in Fig.~\ref{fig1}b]. However, when looking more closely at the energy minima, we find that they display pronounced cusps, not expected from Eq.~\ref{eq:ABS_energy}. These cusps are indicative of phase slips that occur when the superconducting loop has a sizeable inductance,~$L$, whereby the standard linear flux-phase relation no longer holds. We independently estimate $L=\SI{321}{pH}$ (see SI-2) and use the appropriate flux-phase conversion (see SI-6) to find that the measured ABS spectrum is consistent with a large transparency of $\tau=0.99$ (light green line in Fig.~\ref{fig1}b). We further confirm this by performing spectroscopy at higher $B$, as will be discussed later. This highlights the fact that the inductance, which can be significant in thin film superconductors, strongly affects the ABS spectra observed in experiments.

\begin{figure*}[!h]
	\centering
	\includegraphics[width=0.5\textwidth]{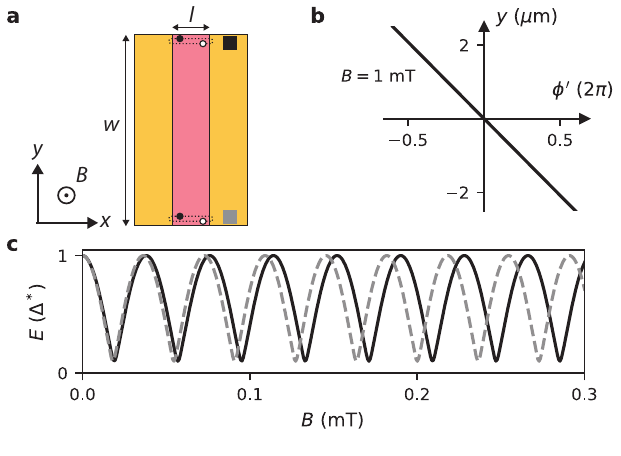}\\
	\caption{Effect of the magnetic vector potential. \textbf{a} Top-view schematic of the JJ in Dev~1. Two ABSs located at the top and bottom end are indicated. \textbf{b} Calculated local phase difference arising from the vector potential at $B=\SI{1}{mT}$ ($f = 6.2$). \textbf{c} Magnetic field evolution for the ABS located at the top (black) and bottom (grey), showing a relative shift due to the local phase difference.}
	\label{fig2}
\end{figure*}

Thus far we have assumed that the superconducting phase difference is constant along the width of the JJ (see Fig.~\ref{fig2}a for a top-view schematic of the junction). However, the vector potential of the magnetic field creates a phase gradient, $\phi^{\prime}(y)$, and the total gauge-invariant phase difference is given by $\varphi(y)=\phi+\phi^{\prime}(y)$, where $\phi$ is the phase difference that can be tuned by the flux through the loop. The position-dependent local phase difference can be expressed as~\cite{Tinkham_1996, Newrock}: 

\begin{equation}
	\phi^{\prime} = -2\pi\frac{fBly}{\Phi_0}, \label{eq:Vec_pot}
\end{equation}

where $f$ is a flux focusing factor that increases the effective magnetic flux in the JJ (see SI-3 and ref.~\onlinecite{Suominen_2017}). This expression for $\varphi$ is valid for JJs with a width much smaller than the Josephson penetration length, which is the case for our junctions (see SI-4). The magnetic vector potential also leads to the formation of localized ABSs with a well defined supercurrent direction (see SI-7 for numerical simulations). Fig.~\ref{fig2}b shows a plot of the expected local phase difference for Dev~1 at $B=\SI{1}{mT}$, demonstrating that the phase difference experienced by an ABS located at the top and bottom end of the JJ will be significantly different. Therefore, for localized ABSs (as depicted in Fig.~\ref{fig2}a), one expects observable differences in the field evolution of their energies. This is more clearly illustrated in Fig.~\ref{fig2}c, where we plot the ABS energy, $E$, as a function of $B$. As $B$ increases, the maxima for the top and bottom ABS shift relative to each other. This is a direct consequence of Eq.~\ref{eq:Vec_pot}, whereby ABSs located at opposite ends of the JJ are sensitive to the local phase difference with equal magnitude but opposite sign.

\begin{figure*}[!t]
	\centering
	\includegraphics[width=0.95\textwidth]{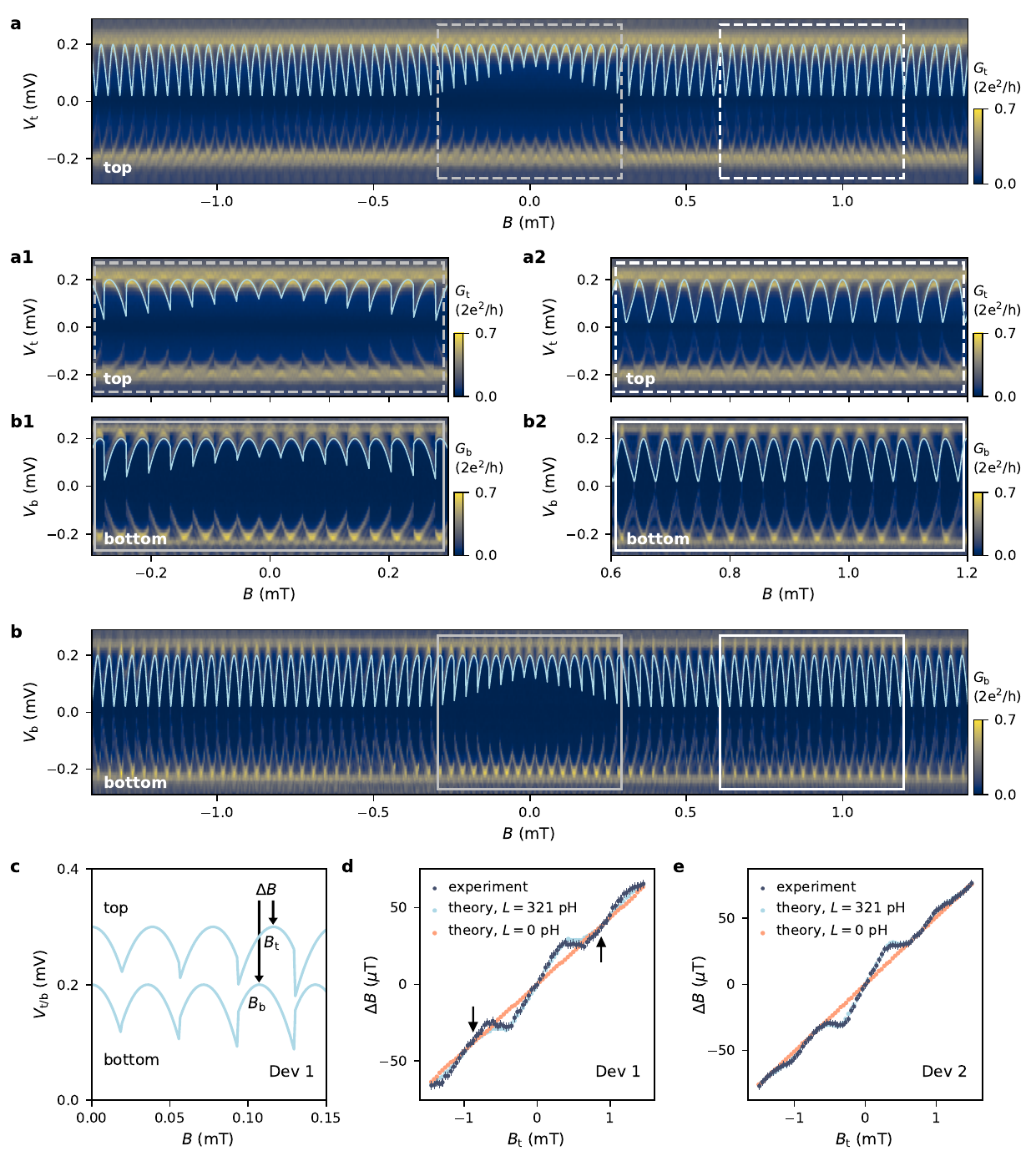}\\
	\caption{Tunneling spectroscopy over a large magnetic field range. \textbf{a} Spectroscopy map at the top end of Dev~1 ($V_{\mathrm{g1}}=\SI{-0.39}{V}$, $V_{\mathrm{g2}}=\SI{-0.74}{V}$, $V_{\mathrm{g3}}=\SI{0}{V}$, $V_{\mathrm{g4}}=\SI{0}{V}$), with zoomed-in views presented in \textbf{a1} and \textbf{a2}. \textbf{b} Spectroscopy map at the bottom end ($V_{\mathrm{g1}}=\SI{0}{V}$, $V_{\mathrm{g2}}=\SI{0}{V}$, $V_{\mathrm{g3}}=\SI{-1.1}{V}$, $V_{\mathrm{g4}}=\SI{-0.6}{V}$) with zoom-in views in \textbf{b1} and \textbf{b2}. The model (light blue lines) assumes coupling to a single ABS ($\tau=0.99$), taking into account the local phase difference in the JJ and the loop inductance ($L=\SI{321}{pH}$). \textbf{c} Model curves for the top and bottom end plotted together (offsetted vertically for clarity). The ABS maxima on the top ($B_{\mathrm{t}}$) and bottom ($B_{\mathrm{t}}$) are shifted. \textbf{d, e} $\Delta B = B_{\mathrm{t}} - B_{\mathrm{b}}$ as a function of $B_{\mathrm{t}}$ for Dev~1 and Dev~2 (dark blue circles). We also include the $\Delta B$ values from the toy model with $L=\SI{321}{pH}$ (light blue circles), and $L=0$ (red circles). The arrows indicate the position of the first Fraunhofer node.}
	\label{fig3}
\end{figure*}

With an understanding of the effect of the magnetic vector potential on the ABS spectrum, we now turn to spectroscopy measurements over a significantly larger field range (Fig.~\ref{fig3}). Figure~\ref{fig3}a and b show the top and bottom spectroscopy maps, respectively. We first look at the high field regime (Fig.~\ref{fig3}a2 and b2), where the ABS oscillation amplitude has increased significantly (compare to~\ref{fig1}b). This is caused by the Fraunhofer-like reduction of the critical current, $I_{\mathrm{c}}$, thereby reducing the so-called screening parameter, $\beta \propto LI_{\mathrm{c}}$. The lower $\beta$ results in a linear flux-phase relation, making it possible to probe the complete phase-dependence of the ABS (see SI-6 for more details). The fact that the ABS energy reaches very close to zero confirms that the ABSs we are probing have extremely high transparency.  

In the intermediate field range (see Fig.~\ref{fig3}a1 and b1) we find that the cusps near the ABS minima develop into sharp jumps, resulting in a highly asymmetric and skewed shape away from $B=0$. The skewness is not only reversed for positive and negative fields, but also for the top and bottom end of the JJ. Furthermore, we find that the ABS energy maxima shift in opposite directions in the top and bottom spectroscopy map, as expected for bound states localized at the edges. This is a strong indication that each probe is sensitive only to a region of limited spatial extent in its vicinity, and that it is in general difficult to reliably estimate bulk junction properties from a local spectroscopy measurement~\cite{Nichele_2020}.

To explain these findings we introduce a model that takes into account the combined effects of the inductance and vector potential, and assumes that each tunnel probe couples only to a single localized ABS with $\tau=0.99$ (a full description of the model can be found in SI-6). The resulting ABS spectra are shown as light blue lines plotted on the spectroscopy maps of Fig.~\ref{fig3}a and b. We find an excellent agreement between the model and the experiments in the entire magnetic field range. We show in SI-6 that the observed reversal of the skewness can only occur when both the vector potential and the loop inductance are taken into account. Therefore, the loop inductance serves as an extremely useful tool to clearly see the effects of a spatially varying phase difference along the JJ.

In order to systematically analyze the difference between the energy spectra of the top and bottom ABS, we introduce the quantity $\Delta B = B_{\mathrm{t}} - B_{\mathrm{b}}$ (see Fig.~\ref{fig3}c). In Fig.~\ref{fig3}d, we plot $\Delta B$ as a function of $B_{\mathrm{t}}$ for experiment (dark blue circles) and theory (light blue circles). Both show a non-linear dependence, which can be well accounted for by the variation of $I_{\mathrm{c}}$ (and hence $\beta$). It is interesting to note that while our device geometry makes it impossible to directly measure $I_{\mathrm{c}}$ of the JJ, the nodes in the Fraunhofer pattern can still be identified by regions where $\beta \approx 0$ (see arrows), and therefore the experiment/theory plots with finite $L$ approach the theory curve with $L = 0$ (red circles). All of these findings are reproduced in Dev~2 (see spectroscopy maps in SI-5 and the $\Delta B$ analysis in Fig.~\ref{fig3}e). 

Although our toy model is effective in capturing the most important features observed in experiments, it relies on the assumption that the tunnel probes couple to a single localized ABS in the vicinity of the barriers. In the following, we use numerical simulations to show that the tunneling current is indeed dominated by edge-located ABSs, and that the phase shifts for these states agree with the experiments. For the simulations, we consider a planar JJ composed of two semi-infinite superconducting leads and a normal region that is connected to two normal leads through tunneling barriers. We calculate the conductance from the top (bottom) normal lead, $G_{t}$ ($G_{b}$), by tracing the quasiparticles entering and leaving the top (bottom) lead. In the simulation, we include the effect of a perpendicular magnetic field and disorder, which results in a finite mean free path, $l_e$. A superconducting phase difference, $\phi$, is imposed between the superconducting terminals (more details about the model can be found in SI-7). 

\begin{figure*}[!t]
	\centering
	\includegraphics[width=1.0\textwidth]{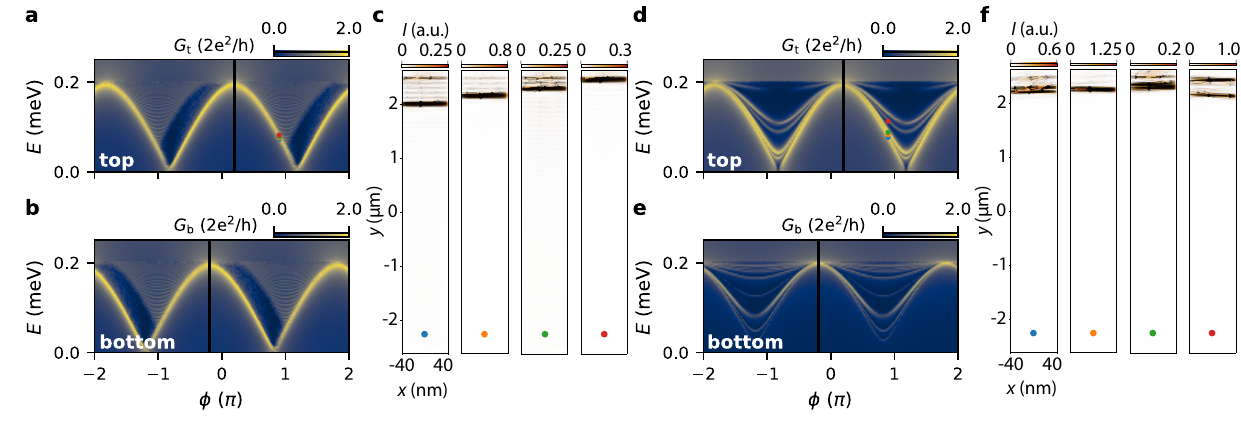}\\
	\caption{Numerical simulation of the tunneling conductance for a ballistic and disordered JJ. \textbf{a, b} Conductance maps at $B =\SI{1}{mT}$ for a ballistic JJ probed from the top and bottom. The black lines correspond to the phase shifts expected from the toy model. \textbf{c} Supercurrent distribution in the normal region of the JJ obtained for the $E$ and $\phi$ values denoted with the circles in \textbf{a}. \textbf{d, e} Conductance maps at $B =\SI{1}{mT}$ for a disordered JJ ($l_e = \SI{150}{nm}$) probed from the top and bottom. \textbf{f} Supercurrent distributions for the $E$ and $\phi$ values denoted with the circles in \textbf{d}.}
	\label{fig4}
\end{figure*}

We first consider a ballistic JJ with infinite mean free path. In Fig.~\ref{fig4}a and b, we show the conductance calculated from the top and bottom, respectively, at $B = \SI{1}{mT}$. In both maps, the main resonance is shifted by an equal amount in $\phi$, but in opposite directions. This shift agrees very well with our toy model (black lines), where we assumed tunnel-coupling to a single ABS localized at the top/bottom end of JJ. The presence of localized ABSs is clearly seen  by inspecting the supercurrent distribution calculated at the energy/phase values denoted by the colored circles in Fig.~\ref{fig4}a. We find that the top probe is only sensitive to the ABSs located in the vicinity of the top barrier (see Fig.~\ref{fig4}c).  

To make a connection with the experiments, we also consider a semiconductor with $l_e=\SI{150}{nm}$, a good estimate for the mean free path in our 2DEGs~\cite{Moehle_2021}. The top and bottom conductances are shown in \ref{fig4}d and e, respectively. As in the ballistic case, we again find a predominant sensitivity to edge-located ABSs, and a relative shift of the ABS maxima. However, we also note two important differences. Firstly, unlike the ballistic case, the ABS spectra at the top and bottom are now drastically different from each other. This is not surprising, given the fact that the ABSs can be sensitive to the particular disorder configuration present at each end. Secondly, the main resonance splits into more clearly distinguishable ABSs. These ABSs are also localized close to the top/bottom end of JJ, as seen in Fig.~\ref{fig4}f. The specific location of these states is sensitive to the local potential landscape. However, we expect them to acquire different relative phase shifts depending on their precise location in the JJ. 

\begin{figure*}[!t]
	\centering
	\includegraphics[width=1.0\textwidth]{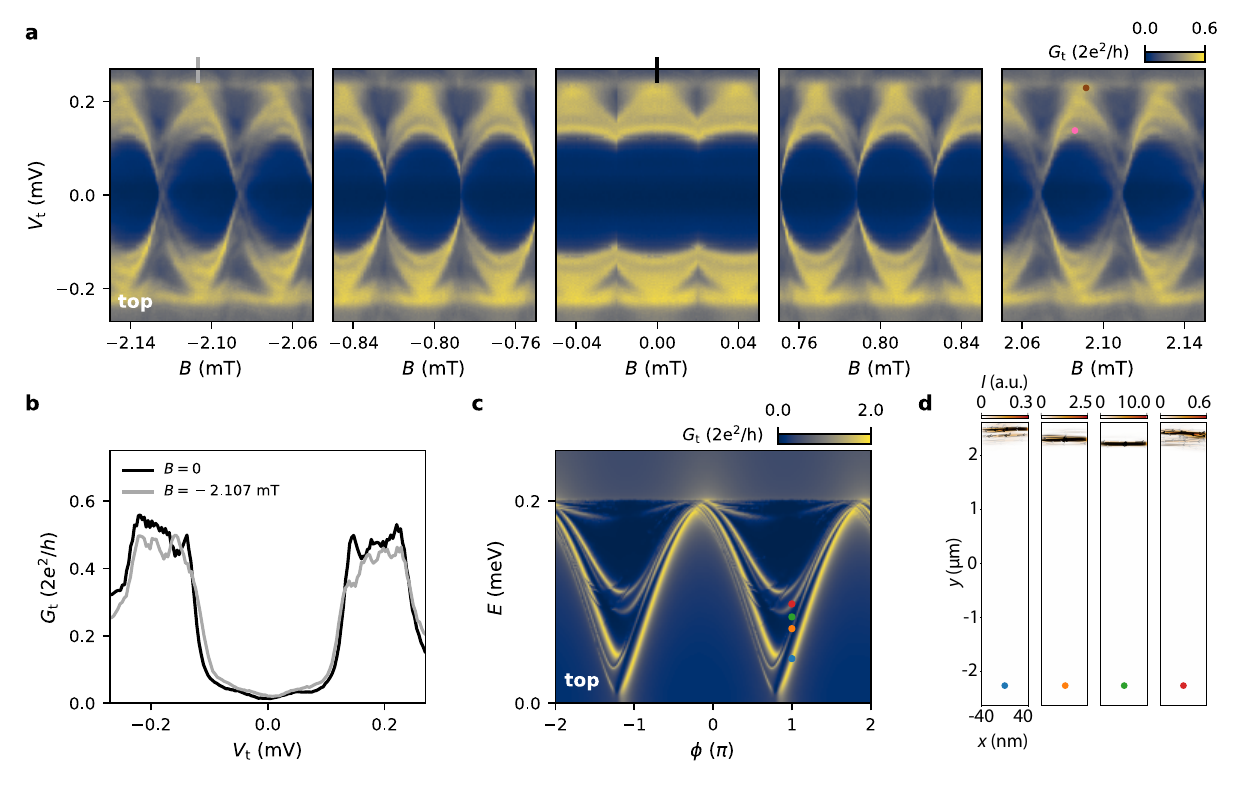}\\
	\caption{Probing spatially separated ABSs. \textbf{a} Tunneling spectroscopy maps at the top end of Dev~2 ($V_{\mathrm{g1}}=\SI{-1.90}{V}$, $V_{\mathrm{g2}}=\SI{-1.40}{V}$, $V_{\mathrm{g3}}=\SI{-2.10}{V}$, $V_{\mathrm{g4}}=\SI{-1.43}{V}$). The ABSs that are initially hardly resolvable around $B=0$ are better resolved at larger $B$, where localized ABSs acquire different phase shifts depending on their location in the JJ. \textbf{b} Line cuts at two indicated positions in \textbf{a} showing this improvement in resolution. \textbf{c} Simulated tunneling conductance map for a disordered JJ ($l_e = \SI{150}{nm}$) at $B = \SI{10}{mT}$ probed from the top. \textbf{d} Supercurrent distributions for the $E$ and $\phi$ values marked by circles in \textbf{c}, showing how localized ABSs at different positions correspond to ABS spectra that are shifted in $\phi$.}
	\label{fig5}
\end{figure*}

This spatially dependent phase shift in the vicinity of the tunnel probe can also be experimentally observed. Figure~\ref{fig5}a presents spectroscopy measurements on the top end of Dev~2, where the split gate settings have been modified to locally alter the disorder landscape. At $B=0$ (central panel), distinct ABSs are hardly visible (see also black line cut in Fig.~\ref{fig5}b). However, when increasing the magnetic field (left and right panel), the localized ABSs acquire different phase shifts making it possible to resolve them more clearly (see also gray line cut in Fig.~\ref{fig5}b). Reversing the field direction leads to ABSs shifted in the opposite direction, as expected for spatially separated ABSs. A similar pattern of ABSs located at different positions close to the edge of the junction and experiencing different phase shifts is obtained in the numerical calculation shown in Fig. \ref{fig5}c and d. This demonstrates that the effect of the vector potential (and resulting local phase difference) can indeed be used to estimate the location of the ABSs in the JJ. Around $B=\SI{2.09}{mT}$, the maxima of the two states (indicated by the brown and pink circles) are shifted by $\approx\SI{5}{\mu T}$. This shift can be translated into an estimate of their spatial separation by using the spectroscopy results at the two extreme ends of the JJ (Fig.~S4 and Fig.~\ref{fig3}e), where we find $\Delta B=\SI{106}{\mu T}$ at $B=\SI{2.09}{mT}$ for ABSs separated by $\SI{5}{\mu m}$. Using this, we can estimate the spatial separation of the two states indicated by the brown and pink circles to be approximately $\SI{250}{nm}$.

In conclusion, we employed local tunneling spectroscopy at two ends of planar phase-biased JJs to study the influence of the magnetic vector potential on the ABS spectrum. The combined effect of inductance and a spatially varying local phase difference results in striking differences in the tunneling spectra measured at the two edges of these junctions. Supporting our experiments with a theoretical toy model and microscopic numerical simulations, we showed that our results are consistent with the measurement of ABSs localized at the ends of the JJ, in the vicinity of the tunnel barriers. Finally, we showed that the effects of the vector potential are not only observable for ABSs separated by microns, but can also be used to estimate the relative locations of ABSs separated by a few hundred nanometers. Our results provide insights into the effects of a spatially varying phase difference on the ABS spectrum in extended JJs, and are relevant for ongoing efforts on investigating topological superconductivity in planar JJs. 

\emph{Additional Note:} During the preparation of this manuscript, we became aware of a related work on tunneling spectroscopy in planar JJs~\cite{Banerjee_2022_phase_texture}.\\

\noindent
\textbf{\large Associated content}\\
Device fabrication, estimation of loop inductance, flux focusing in planar JJ, tunneling spectroscopy for Dev~2, toy model, microscopic model\\

\noindent
\textbf{\large Data availability}\\
Raw data and analysis scripts for all presented figures are available at the 4TU.ResearchData repository: https://doi.org/10.4121/20059364.\\

\noindent
\textbf{\large Author contributions}\\
{C.M.M. and C.T.K. fabricated the devices. C.M.M., P.K.R. and N.A.J. performed the measurements and analyzed the data. S.G. supervised the experimental work. The numerical simulations were performed by D.K. under the supervision of M.P.N. who also provided the toy model. The semiconductor heterostructure was grown by D.X. and C.T. under the supervision of M.J.M. The manuscript was written by C.M.M., P.K.R., D.K., M.P.N. and S.G. with input from all authors.}\\

\noindent
\textbf{\large Notes}\\
The authors declare no competing interests.\\

\noindent
\textbf{\large Acknowledgments}\\
We thank T. Dvir, G. Wang and  C. Prosko for fruitful discussions. The research at Delft was supported by the Dutch National Science Foundation (NWO), the Early Research Programme of the Netherlands Organisation for Applied Scientific Research (TNO) and a TKI grant of the Dutch Topsectoren Program. The work at Purdue was funded by Microsoft Quantum. The research at Krakow was supported by National Science Centre (NCN) agreement number UMO-2020/38/E/ST3/00418.\\

\clearpage
\newcommand{\beginsupplement}{%
	\setcounter{section}{0}
	\renewcommand{\thesection}{\arabic{section}}%
	\setcounter{table}{0}
	\renewcommand{\thetable}{S\arabic{table}}%
	\setcounter{figure}{0}
	\renewcommand{\thefigure}{S\arabic{figure}}%
	\setcounter{equation}{0}
	\renewcommand{\theequation}{S\arabic{equation}}%
}

\onecolumngrid

\section{Supplementary Information}
\beginsupplement

\section{1. Device Fabrication}

The two phase-biased JJs (Dev~1, Dev~2) and the DC SQUID are fabricated using electron beam lithography. Due to a possible intermixing of Al and Sb we perform all fabrication steps at room temperature unless stated otherwise. The device fabrication starts by etching the Al and the 2DEG in unwanted areas. The Al etch is performed in Transene~D etchant at a temperature of \SI{48.2}{\degreeCelsius} for 9 s resulting a etching thickness of 100~nm. Afterwards, using the same PMMA mask, the 2DEG is etched in a solution of 560~ml deionized water, 9.6~g citric acid powder, 5~ml $\mathrm{H}_{2}\mathrm{O}_{2}$ and 4~ml $\mathrm{H}_{3}\mathrm{PO}_{4}$, using an etching time of 90~s. To define the JJs, we perform a second Al etch, carried out in \SI{38.2}{\degreeCelsius} Transene~D for 16~s. This is followed by sputtering a 60~nm thick layer of $\mathrm{SiN}_x$ that partly covers the superconducting loop, isolating it from the intended 2DEG contact inside the loop. Next, we contact the exposed 2DEG region on the top and bottom side of the JJ by Ti/Au. Prior to the evaporation of 10~nm Ti and 190~nm Au, a gentle Ar etching is performed in the loadlock of the evaporator to remove any oxides that might have formed on the 2DEG. Afterwards, we contact the superconducting loop by sputtering 150~nm of NbTiN (before the sputter process an in-situ Ar etch is performed to remove the oxide on the Al).
As the gate dielectric, we deposit a global layer (40~nm thick) of $\mathrm{AlO}_x$ by atomic layer deposition at \SI{40}{\degreeCelsius}. The gates are formed in two steps: First, the fine structures (split gates and central gate) are deposited by evaporating 10~nm of Ti and 40~nm of Au. In the second step, 10~nm Ti and 100~nm Au are evaporated to define the gate leads.\\

\begin{figure*}[!h]
	\centering
	\includegraphics[width=0.75\textwidth]{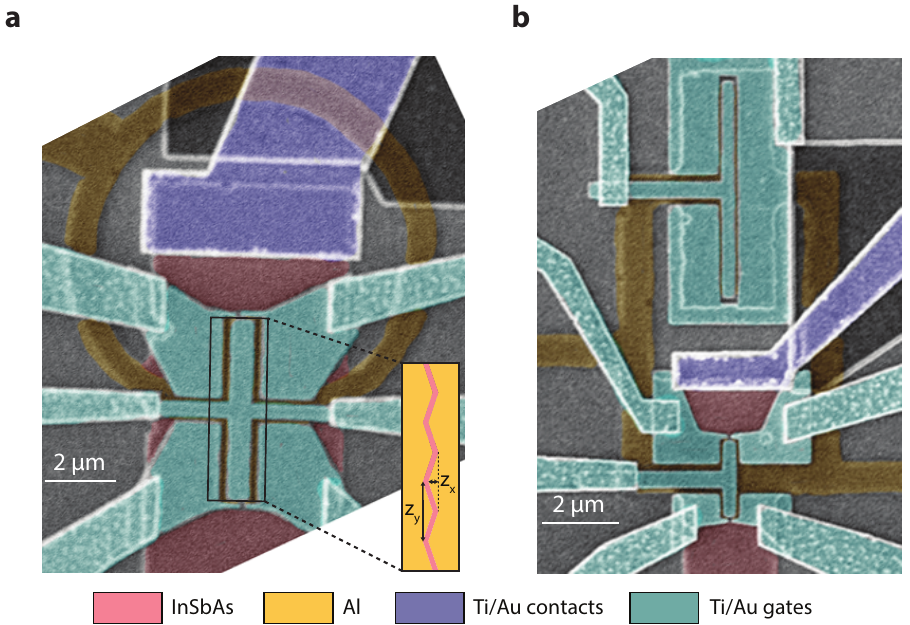}\\
	\caption{\textbf{a} SEM of Dev~2 having a zigzag-shaped normal region ($z_x=\SI{0.24}{\mu m}$, $z_y=\SI{1.43}{\mu m}$) with a length of $l=\SI{80}{nm}$ and width of $w=\SI{5}{\mu m}$. \textbf{b} SEM of the DC SQUID. The device JJ (on the bottom) has dimensions $l=\SI{120}{nm}$ and $w=\SI{2}{\mu m}$. The reference JJ (on the top) has dimensions $l=\SI{80}{nm}$ and $w=\SI{5}{\mu m}$.}
	\label{supp_devices}
\end{figure*}

A schematic and false-colored SEM of Dev~1 is shown in Fig. 1a of the main text. In Fig. \ref{supp_devices}a we present a SEM of Dev~2, which is similar to Dev~1. The main difference is that the normal region of the JJ is slightly zigzag-shaped ($z_x=\SI{0.24}{\mu m}$, $z_y=\SI{1.43}{\mu m}$). This was originally introduced into this device to potentially suppress long quasiparticle trajectories and thereby increase the size of the topological gap~\cite{Laeven_2020}. The superconducting leads of Dev~1 and Dev~2 have a length of 500 nm. Figure \ref{supp_devices}b shows a SEM of the DC SQUID, consisting of two JJs (device JJ and reference JJ) in the superconducting loop. The device JJ has a superconducting lead length of 300 nm. Two additional gates are deposited, one covering the normal region of the reference JJ and one covering the 2DEG region around this junction (always kept at -2.5 V to deplete the 2DEG there).\\

\section{2. Estimation of loop inductance}
\begin{figure*}[!h]
	\centering
	\includegraphics[width=0.7\textwidth]{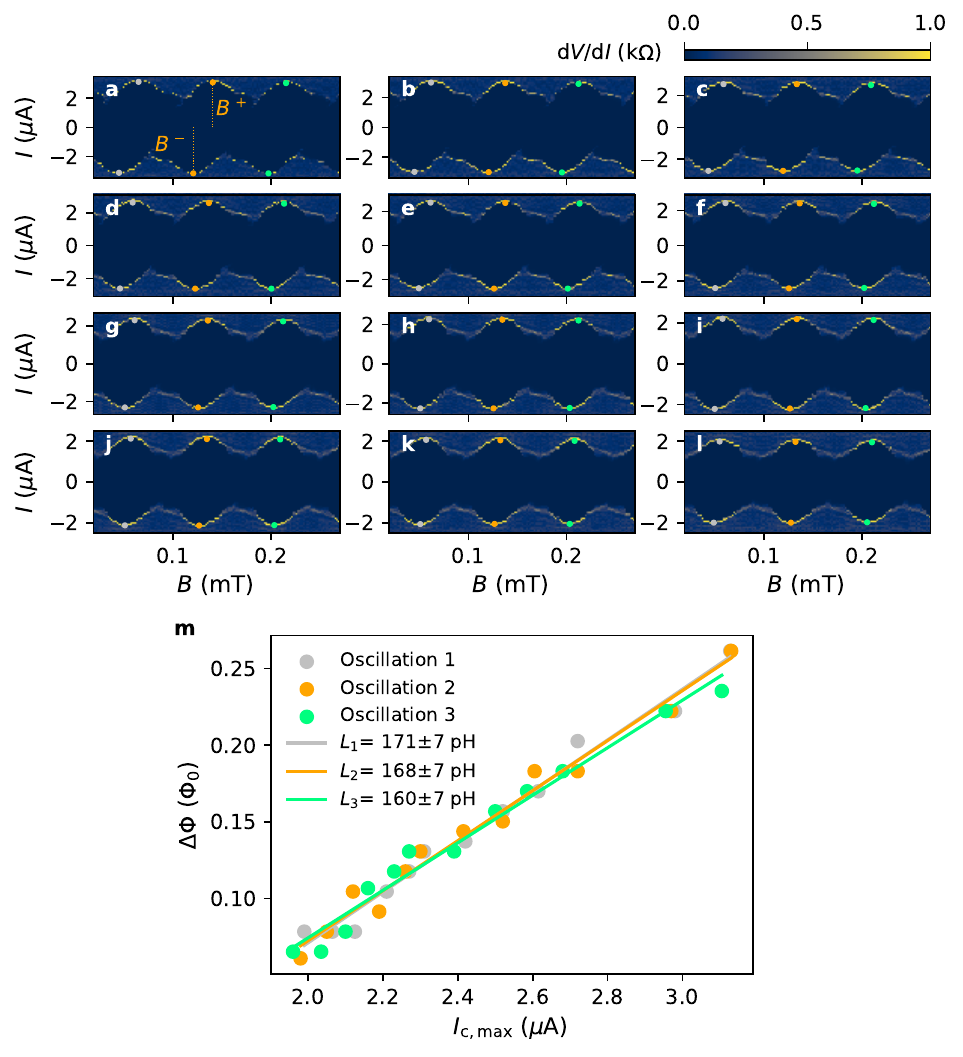}\\
	\caption{Differential resistance, $\mathrm{d}V/\mathrm{d}I$, as a function of applied current bias, $I$, and perpendicular magnetic field, $B$. Panel \textbf{a}-\textbf{l} corresponds to reference gate voltage $V_{\mathrm{g,ref}} =$ 0, -0.4, -0.8, -0.9, -1, -1.1, -1.2, -1.25, -1.3, -1.35, -1.4, and -1.45 V, respectively. No voltage is applied to the device junction gate. The colored circles mark the positions of maximum total critical current. \textbf{m} $\Delta \Phi$ plotted against $I_{\mathrm{c,max}}$ for the three oscillations shown in \textbf{a}-\textbf{l}. The extracted $\Delta B$ is normalized with respect to the oscillation period, giving $\Delta \Phi$ in units of the magnetic flux quantum, $\Phi_0$. The average value of the maximum critical current on the positive and negative current bias sides gives $I_{\mathrm{c,max}}$.}
	\label{supp_inductance}
\end{figure*}

In order to extract the inductance of the SQUID loop, we measure the SQUID interference pattern for different reference JJ gate voltages, $V_{\mathrm{g,ref}}$. Figure \ref{supp_inductance}a-l shows the obtained differential resistance maps as a function of applied current bias, $I$, and perpendicular magnetic field, $B$. Panel a-l corresponds to $V_{\mathrm{g,ref}}=$ 0, -0.4, -0.8, -0.9, -1, -1.1, -1.2, -1.25, -1.3, -1.35, -1.4, and -1.45 V, respectively. The device JJ gate is grounded in all measurements. With the colored circles we mark the positions where the total critical current is maximum. For a given SQUID oscillation, the field at which the maximum occurs is different for positive and negative current bias: $\Delta B = B^+ - B^-$. The corresponding flux difference is given by: $\Delta \Phi = 2(L_{\mathrm{ref}} I_{\mathrm{c,ref}} - L_{\mathrm{dev}} I_{\mathrm{c,dev}})$~\cite{Clarke_2004}.
Here, $I_{\mathrm{c,ref}}$ and $I_{\mathrm{c,dev}}$ are the critical current of the reference and device junction, respectively. The inductances of the two SQUID arms are $L_{\mathrm{ref}}$ and $L_{\mathrm{dev}}$. The above expression can be rewritten as: $\Delta \Phi = 2L_{\mathrm{ref}} I_{\mathrm{c,max}} - 2LI_{\mathrm{c,dev}}$, using the relations for the maximum critical current, $I_{\mathrm{c,max}} = I_{\mathrm{c,ref}} + I_{\mathrm{c,dev}}$, and the total loop inductance, $L = L_{\mathrm{ref}} +L_{\mathrm{dev}}$.\\

In Fig. \ref{supp_inductance}m we plot the extracted $\Delta \Phi$ as a function of $I_{\mathrm{c,max}}$ for the three oscillations indicated in Fig. \ref{supp_inductance}a-l. The linear fits yield $L_{\mathrm{ref}}=166$ pH as the average value. Since the width and the thickness of the superconducting loop is the same for all three devices, the inductance should only depend on the length of the superconducting loop. Under this assumption the loop inductance of the phase-biased JJs can be estimated to be $L_{\mathrm{ref}} l_{\mathrm{PBJJ}}/l_{\mathrm{ref}} = \SI{321}{pH}$, where $l_{\mathrm{ref}} = \SI{15.3}{\mu m}$ is the length of SQUID reference arm and $l_{\mathrm{PBJJ}} = \SI{29.6}{\mu m}$ is the loop length of Dev~1 and~2.

\section{3. Flux focusing in planar JJ}
\begin{figure*}[!h]
	\centering
	\includegraphics[width=0.6\textwidth]{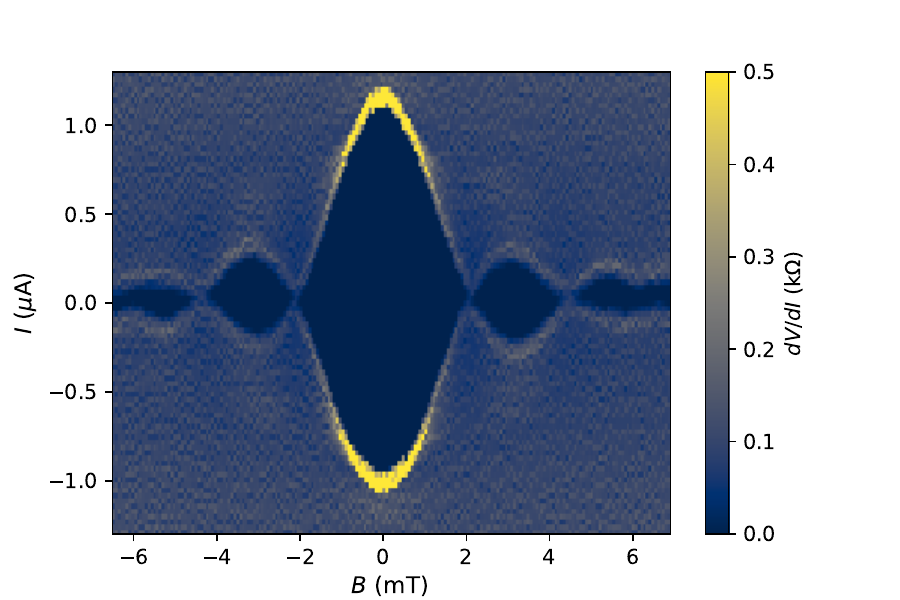}\\
	\caption{Differential resistance, $\mathrm{d}V/\mathrm{d}I$, as a function of applied current, $I$, and perpendicular magnetic field, $B$ for the device JJ of the DC SQUID.}
	\label{supp_flux_focusing}
\end{figure*}

The Fraunhofer interference pattern periodicity, $B_\mathrm{0}$, in a JJ is determined by the geometrical area, $A$, enclosed between two superconducting leads, i.e. $B_\mathrm{0} = \Phi_\mathrm{0}/ A$. However, in the presence of flux focusing, the periodicity is reduced from the theoretical value~\cite{Suominen2017}. To estimate the effect of flux focusing we measure the differential resistance, $\mathrm{d}V/\mathrm{d}I$, as a function of applied current, $I$, and perpendicular magnetic field, $B$, for the device JJ of the DC SQUID (see Fig.~\ref{supp_flux_focusing}). For this measurement, the reference JJ is pinched off by applying a voltage of $\SI{-2.5}{V}$ to the top gate. We observe the first node at $\SI{2.1}{mT}$ instead of the expected Fraunhofer periodicity of $B_\mathrm{0} = \SI{8.6}{mT}$. This gives a dimensionless flux focusing factor, $f$, of 4.1 for this junction.

To explain our spectroscopy maps measured at the top and bottom ends of Dev~1 and Dev~2 we introduce a toy model with flux focusing in Sec.~\ref{section: Toy model}. Although the above extracted $f$ gives an estimate of the focusing factor, the exact value can vary from junction to junction. The best agreement between the experimental spectroscopy maps and the toy model is be achieved with $f = 6.2$ for Dev~1 and $7.2$ for Dev~2 (see Fig.~3 in the main text as well as Fig.~\ref{supp_fig_zigzag}]. The larger $f$ values (and therefore stronger flux focusing) are in fact expected due to the shorter JJ length and larger lead length of Dev~1 and Dev~2 compared to the values for the device JJ of the DC SQUID~\cite{Suominen2017}.

\section{4. Josephson penetration depth}

The Josephson penetration depth for a JJ with the thickness of the superconducting electrodes comparable or smaller than the penetration depth is dominated by the kinetic inductance contribution and is given as~\cite{tolpygo_1996_critical}: $\lambda_J = (\Phi_0 w / 4\pi \mu_0 J_c \lambda^{2} )^{1/2}$, where $w=\SI{5}{\mu m}$ is the junction width, $J_c$ is the critical current density, and $\lambda$ is the superconducting penetration depth of Al. 

For our junctions, the thickness of the Al electrodes (7 nm) is much smaller compared to the previously reported value of $\lambda =$~180 nm for a similar heterostructure~\cite{Suominen2017}. Therefore we use the above expression to determine $\lambda_J$. Since the critical current cannot be measured for Dev~1 and Dev~2, we estimate it based on values obtained for the DC SQUID. The critical current of the device JJ with width $w = \SI{2}{\mu m}$ is $I_c = \SI{1.05}{\mu A}$ (see Fig.~\ref{supp_flux_focusing}) and the thickness of 2DEG is $t=\SI{30}{nm}$. Using these values we get $J_c = I_c/wt = \SI{1.75} {\times 10^7 A/m^{2}}$ and $\lambda_J = \SI{34}{\mu m}$, which is much larger than the width of the JJs ($w=\SI{5}{\mu m}$). This ensures that the gauge-invariant phase difference can be expressed as $\varphi(y)=\phi+\phi^{\prime}$, with $\phi^{\prime} = -2\pi\frac{fBly}{\Phi_0}$.

\section{5. Tunneling spectroscopy for Dev~2}

\begin{figure*}[!h]
	\centering
	\includegraphics[width=1.0\textwidth]{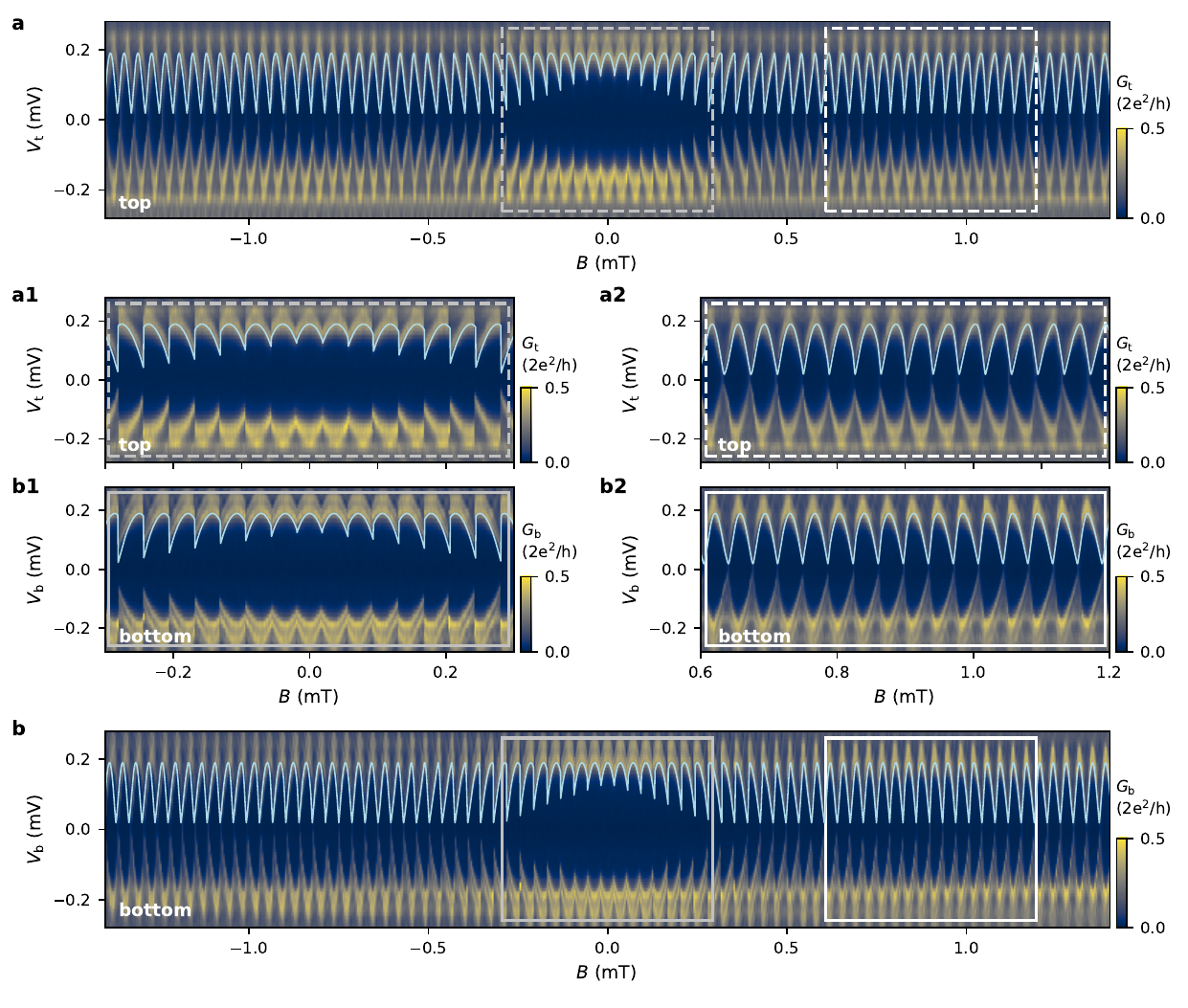}\\
	\caption{\textbf{a} Spectroscopy map at the top end of Dev~2 with zoom-ins presented in \textbf{a1} and \textbf{a2}. The bottom spectroscopy map is shown in \textbf{b} with zoom-ins in \textbf{b1} and \textbf{b2}. Both measurements were obtained with $V_{\mathrm{g1}}=\SI{-1.60}{V}$, $V_{\mathrm{g2}}=\SI{-1.42}{V}$, $V_{\mathrm{g3}}=\SI{-2.10}{V}$, $V_{\mathrm{g4}}=\SI{-1.43}{V}$. The model (light blue line) assumes coupling to a single ABS ($\tau=0.99$), taking into account the local phase difference in the JJ and the loop inductance ($L=\SI{321}{pH}$) for the field-phase conversion.}
	\label{supp_fig_zigzag}
\end{figure*}

\clearpage
\section{6. Toy model}
\label{section: Toy model}
This model is used to calculate the Andreev bound states (ABSs) energies of a Josephson junction embedded in a superconducting loop in the presence of a perpendicular magnetic field, as used to substantiate the measurement results shown in Fig.~3 and Fig.~S5. The junction has a length $l$ (the distance between the superconducting contacts) and a width $w$ (the distance between the edges of the junction where the tunneling probes are connected).

We assume a homogeneous density of the supercurrent in the junction and that the current is carried by $M$ ABSs uniformly distributed across the junction at positions $y_n = -w/2 + (n-1)\cdot w/(M-1)$ with integer $n \in [1,M]$.

The positive energies of the ABSs in the junction with the transmission coefficient $\tau$ are given by~\cite{PhysRevLett.67.3836}:
\begin{equation}
E_n(\varphi_n) = \Delta \sqrt{1- \tau \sin^2(\frac{\varphi_n}{2})},
\label{ABS_energy_supp}
\end{equation}
where, in the presence of the external perpendicular magnetic field, $\varphi_n = \phi + \frac{2 \pi}{\Phi_0} \int_{(0,y_n)}^{(l,y_n)} \mathbf{A}\cdot d\mathbf{l}$ is the gauge-invariant phase drop across the junction for an ABS located at position $y_n$. $\phi$ is the superconducting phase difference. For the vector potential in the Landau gauge $\mathbf{A} = (-yB, 0, 0)$, the phase drop in the junction at $y_n$ is $\varphi_n = \phi -(2 \pi/\Phi_0) \cdot f B l y_n$, where we included $f$ as the magnetic field focusing factor. The latter equation gives the phase evolution of the ABS located at the edges of the junction as $\varphi_{\mathrm{t/b}} = \phi \mp (\pi/\Phi_0) \cdot f B l w$~\cite{PhysRevB.102.165407} with a minus (plus) sign for the upper (bottom) edge.
 
The zero-temperature supercurrent obtained from the positive-energy ABSs in the junction is given by:
\begin{equation}
I(\varphi) = \frac{e \Delta^2 \tau}{2\hbar}\sum_n^M\frac{\sin(\varphi_n)}{E_n(\varphi_n)}.
\end{equation}

\begin{figure}[!h]
\center
\includegraphics[width = 14cm]{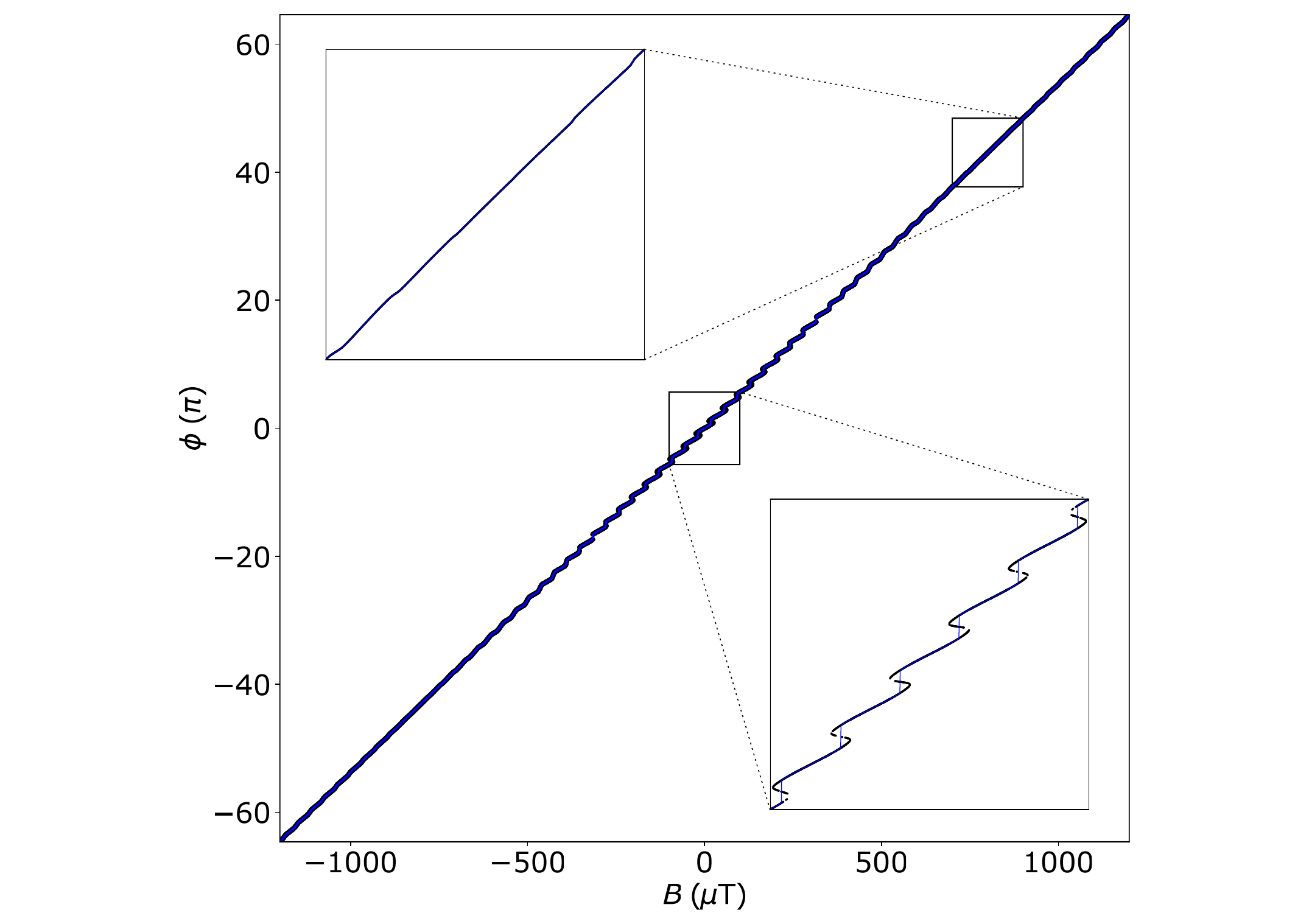}
\caption{Superconducting phase difference versus applied magnetic field obtained for $L = 321$~pH, $l = 80$~nm, $w = 5000$~nm, $M = 35$, $\tau = 0.99$, $R = 4207$~nm, $\Delta = 0.2$~meV and $f = 6.2$. The black dots show possible phase values for a given $B$, while the blue curve shows the superconducting phase difference obtained by minimizing the total energy.}
\label{supp_figure_flux_to_phase}
\end{figure}

In the experimental setup, the superconducting phase difference $\phi$ is induced by a flux $\Phi = B \pi R^2$ that threads a superconducting loop with radius $R$. The non-zero loop inductance $L$ results in the following phase-flux relation~\cite{Banerjee2022}:
\begin{equation}
    \phi = \frac{2 \pi}{\Phi_0}(\Phi - L I(\varphi)).
\label{fluxphase}
\end{equation}

We obtain the energies of the ABSs located at the edges of the junction versus $B$ using the following procedure. In the first step, we solve Eq.~\ref{fluxphase} for a given $B$ and obtain the $\phi$ value that minimizes the total energy of the system $E(\phi) = LI^2(\varphi)/2 - \sum_n^M E_n(\varphi_n) $ calculated as the sum of the energy contained in the superconducting loop and the free energy of the junction ($F = const - E_j = const - \sum E_n$). An example of a flux-to-phase conversion curve is shown in Fig.~\ref{supp_figure_flux_to_phase}. Finally,  we use the phase difference value to calculate $E_n$ corresponding to the ABSs located at the outermost edges of the junction using Eq.~\ref{ABS_energy_supp}. 

\begin{figure}[!t]
\center
\includegraphics[width=1.0\textwidth]{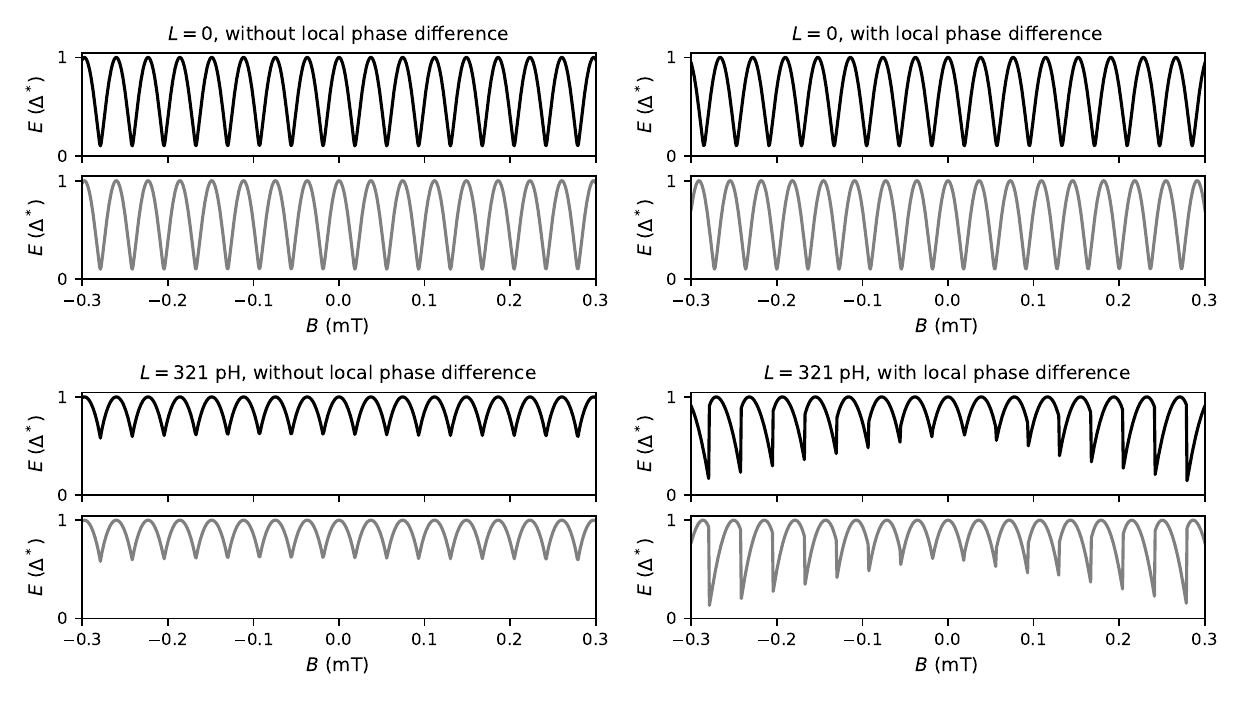}
\caption{ABS energy versus magnetic field in the presence or absence of the loop inductance and with or without local phase difference in the JJ as indicated. The top and bottom panels correspond to an ABS located at the top and bottom end of the JJ, respectively. For all plots, the following parameters are used: $l = 80$~nm, $w = 5000$~nm, $M = 35$, $\tau = 0.99$, $R = 4207$~nm, $\Delta = 0.2$~meV and $f = 6.2$.}
\label{supp_figure_L_vecpotential}
\end{figure}

Figure~\ref{supp_figure_L_vecpotential} shows an ABS located at the top (top panel) and bottom (bottom panel) end of the JJ in the presence and absence of the loop inductance and the local phase difference arising from the magnetic vector potential as indicated. The reversal of the skewness can only happen when both the loop inductance and the local phase difference are present.

Table~\ref{Tab_Parameters} summarizes the parameters that are used for the overlays for Dev~1 (Fig.~3 of the main text) and Dev~2 (Fig.~\ref{supp_fig_zigzag}).

\begin{table}[!h]
\centering
\begin{tabular}{  c  | c  | c }
\hline
Parameter & Dev 1 & Dev 2 \\
\hline
$l$ (nm) & 80 & 80 \\
$w$ ($\mu$m) & 5 & 5\\
$R$ (nm) & 4207 & 4190 \\
$L$ (pH) & 321 & 321 \\
$\Delta$ (meV) & 0.2 & 0.19 \\
$f$ & 6.2 & 7.2\\
$M$ & 35 & 45\\
$\tau$ & 0.99 & 0.99\\
\hline
\end{tabular}
\caption{Toy model parameters used for Dev~1 and Dev~2.}
\label{Tab_Parameters}
\end{table}

\clearpage

\section{7. Microscopic model}
\subsection{A. Tunneling spectroscopy calculations}
\begin{figure}[!t]
\center
\includegraphics[height = 8cm, width = 3.2cm]{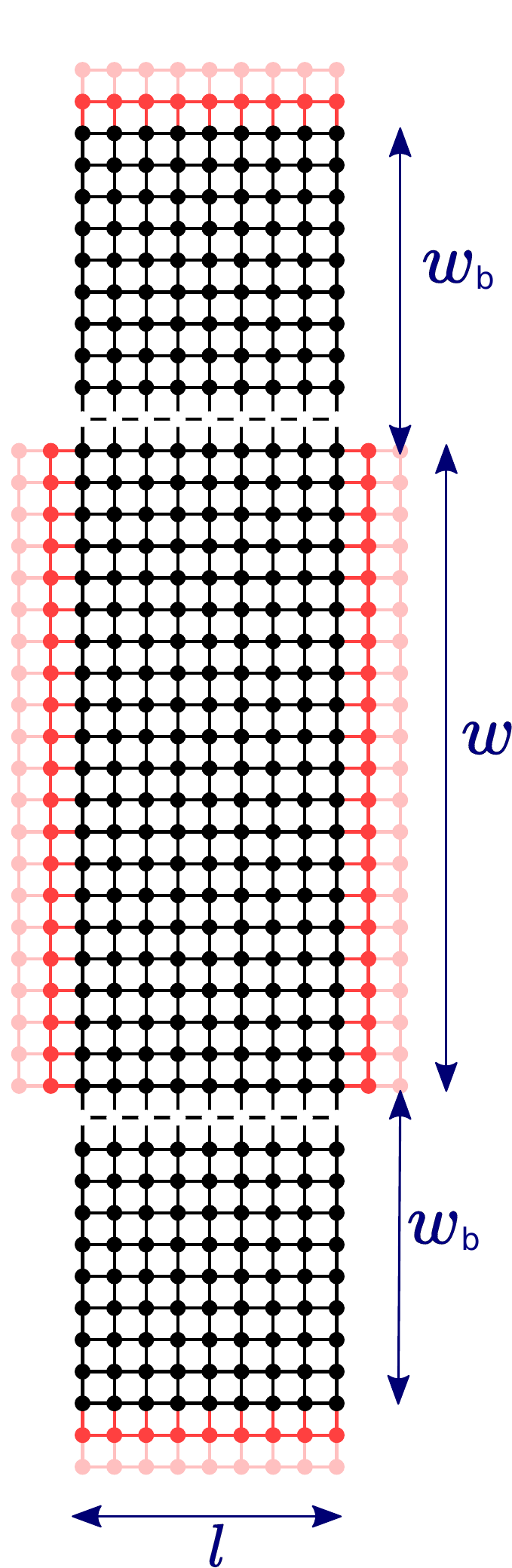}
\caption{Schematic of the system considered for tunneling spectroscopy calculations. The dots denote the sites of the computational mesh. The black dots correspond to the scattering region, whereas the pink ones denote the semi-infinite leads. We use $l$ = 80 nm (the distance between the superconducting contacts) and $w = 5000$ nm (the distance between the edges of the junction where the tunneling probes are connected). The barrier potential at the top and bottom is separated from the normal leads of width $w_{b}=100$ nm. The vertical leads are superconducting, while the horizontal leads are normal.\\
}
\label{system}
\end{figure}

We consider a four-terminal device, with two vertical superconducting leads separated by the normal region (which creates the superconductor-normal-superconductor junction) and two normal leads that are placed horizontally at the top and bottom---see Fig.~\ref{system}. Between the horizontal leads and the normal scattering region, we introduce tunneling barriers that mimic the behavior of QPCs tuned into the tunneling regime.

The considered system is described by the Hamiltonian
\begin{equation}\label{eq:1}
 \left[ \begin{array}{cc} H & \Delta \\ \Delta^* & -H  \\ \end{array} \right],
\end{equation} \\
acting on a wave function in the basis $\Psi = (\Psi_e, \Psi_h)^T$. Here $H$ is defined as 
\begin{equation}\label{eq:2}
    H = -\frac{\hbar^2}{2m^*} \nabla^2 + V(r) - \mu.
\end{equation}
$\mu$ is the chemical potential, $m^*$ is the effective electron mass and $V(r)$ is the rectangular potential barrier of height $V_{g}$ placed just above and below the normal region of length ($l =80$ nm).  

In the presence of a magnetic field, the Hamiltonian $H$ becomes 
\begin{equation}\label{eq:3}
    H^{'} = -\frac{\hbar^2}{2m^*} (\nabla-q\mathbf{A}/\hbar)^2 + V(r) - \mu, 
\end{equation}
with $q = -|e|$ for the electron and $q = |e|$ for the hole part of the Hamiltonian Eq. \ref{eq:1}. We choose the vector potential in the Landau gauge with $\vec B = B\hat{z}$, so that $\vec A = - By\hat{x}$

The superconducting pairing potential $\Delta$ varies spatially and is given by:
\begin{equation*}
\Delta(x) = \begin{cases*}
                    \Delta_{0} & if  $x < -l/2$  \\
                     \phantom{}0 & if $-l/2 \le x \le l/2$\\
                     \phantom{}\Delta_{0}e^{\iota\phi} & if $x > l/2$
                 \end{cases*} 
\end{equation*}

We discretize the Hamiltonian Eq.~\ref{eq:1} on a square lattice with discretization constant $a = 10$ nm. We put the material parameters as $m^* = 0.016m_{e}$, $\mu = 5$ meV, $\Delta = 0.2$ meV. We introduce the anisotropic mass in the superconducting leads with the mass along the translation symmetry of the superconducting leads equal to $10m^*$ as appropriate for the description of transparent normal-superconductor interfaces in models where the chemical potential is kept constant~\cite{short_junctions}. Including a vector potential in this system is done using Peierls substitution as $t_{nm} \to t_{nm}\exp[\frac{-\iota e}{\hbar}\int \mathbf{A}d\mathbf{l}]$~\cite{Peierls1933,PhysRevB.14.2239}.

We exclude the magnetic field from the superconducting leads to account for the screening effect setting $\mathbf{A}=0$ there. We also put zero vector potential in the top and bottom leads to maintain the translation invariance. This in turn introduces a delta peak in the magnetic field where the horizontal leads are attached (as calculated from $\mathbf{B} = \nabla\times{\mathbf{A}}$). We have, however, verified that for the considered small magnetic fields, this does not affect our results, as confirmed by replacing the vector potential by a position-dependent superconducting phase as
$\phi \to \phi - \frac{2\pi Bly}{\Phi_{0}}$ and observing that both results match accurately.

The finite mean free path ($l_e$) is implemented by introducing a random on-site disorder potential $V_d(x,y)$ with the amplitude uniformly distributed between $-U_d/2$ and $U_d/2$~\cite{disorder}, where
\begin{equation}\label{eq:5}
 U_{d} = \mu \sqrt{\frac{6{\lambda_{F}^3}}{\pi^3 a^2 l_e}}.  
\end{equation} 
Here $a, l_{e}, \lambda_{F}$ are the lattice constant, mean free path and the Fermi wavelength, respectively.
We calculate the conductance map with respect to the phase difference $\phi$ and energy using the scattering matrix approach implemented in the Kwant package~\cite{kwant}, using the formula:
  \begin{equation}\label{eq:4}
     G_{t/b} = \frac{2e^2}{h}(N_{t/b}^e - T_{t/b}^{ee} + T_{t/b}^{he}),
 \end{equation}
where $t$ and $b$ stand for top and bottom lead respectively and $N_{t/b}^e$ is the corresponding number of the electron modes.
 The energy dependent transmissions are evaluated as: 
 \begin{equation}
   T_{t/b}^{\alpha\beta} = Tr([S_{t/b}^{\alpha\beta}]^\dagger S_{t/b}^{\alpha\beta}),
 \end{equation}
 where $S_{t/b}^{\alpha\beta}$ is the block of scattering amplitudes of incident particle of type $\beta$ in $t$ ($b$) lead to a particle of type $\alpha$ in the lead $t$ ($b$).
\begin{figure}[ht!]
\center
\includegraphics[width = 1.0\textwidth]{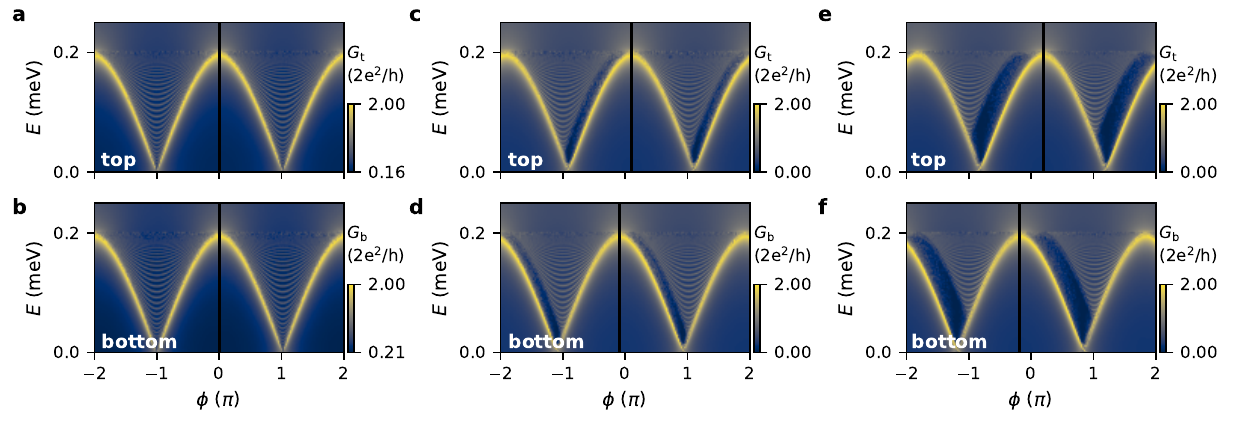}
\caption{Conductance versus phase difference and energy calculated for quasiparticles injected from the top lead (upper row) and bottom lead (lower row) at $B = 0$ (\textbf{a},\textbf{b}), $B = 0.5$ mT (\textbf{c},\textbf{d}) and $B = 1$ mT (\textbf{e},\textbf{f}). The vertical black lines denote the expected phase shift of the edge modes due to the magnetic field $\varphi_{\mathrm{t/b}} = \phi \mp (\pi/\Phi_0) \cdot f B l w$.}
\label{supp_figure_varied_B}
\end{figure}

\begin{figure}[h!]
\center
\includegraphics[width = 1.0\textwidth]{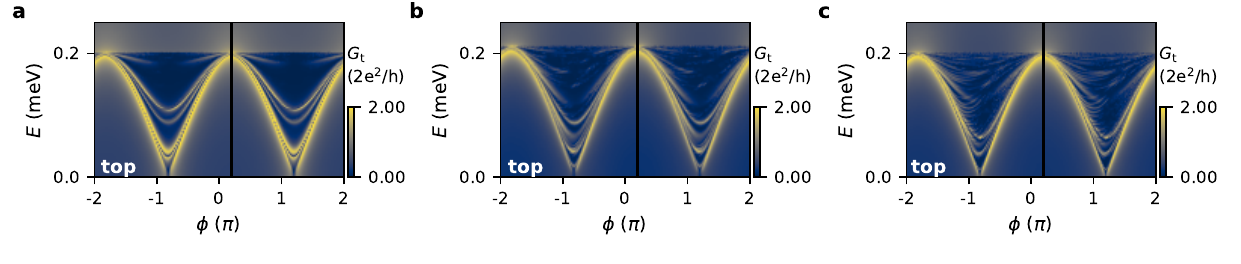}
\caption{Conductance versus phase difference and energy calculated for quasiparticles injected from the top lead for $B = 1 $ mT and different mean free paths $l_e =$ 150 \textbf{a}, 500 \textbf{b} and 1000 nm \textbf{c}.}
\label{supp_figure_varied_le}
\end{figure}
 
\subsection{B. ABS calculation}
\begin{figure}[!h]
\center
\includegraphics[width = 0.9\textwidth]{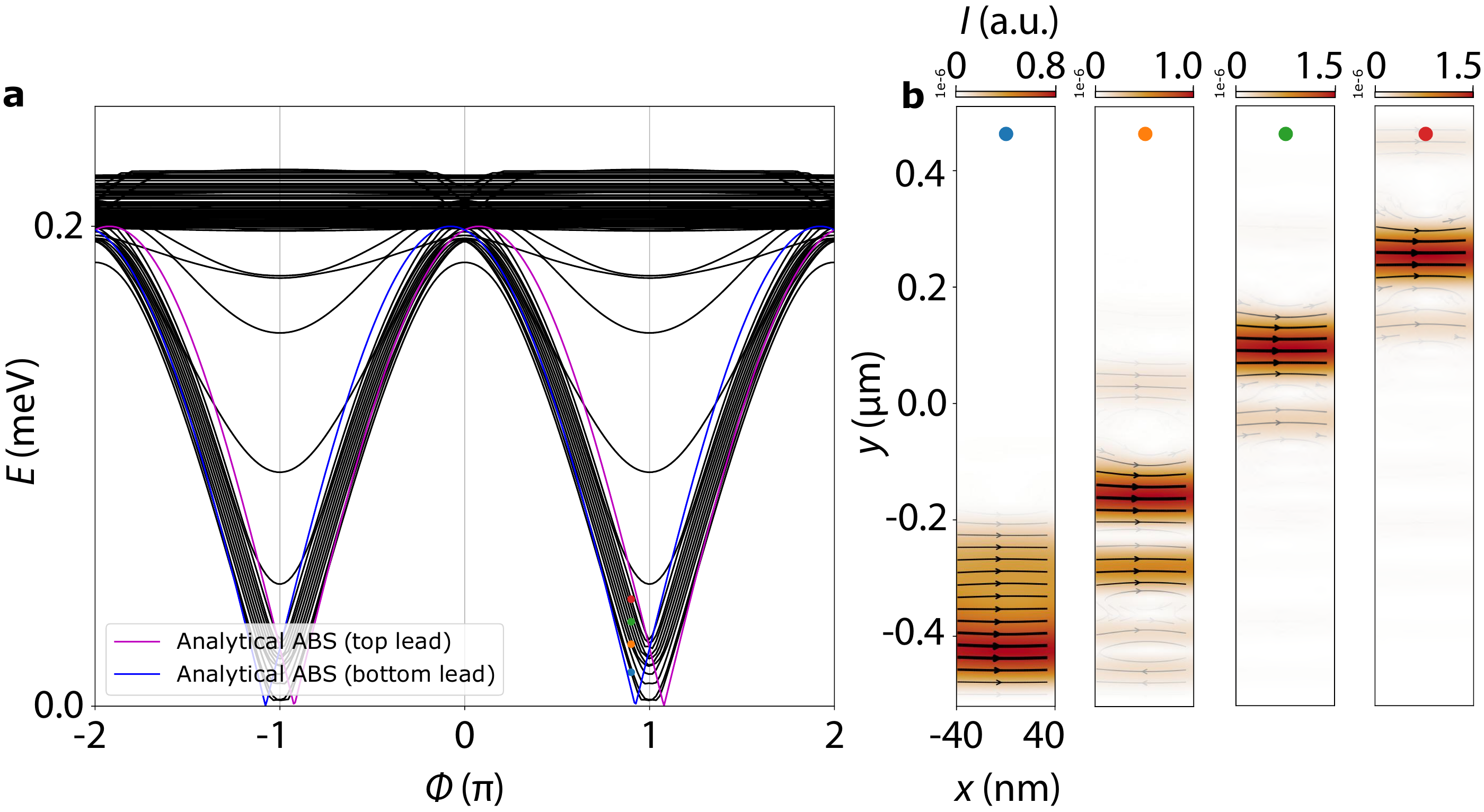}
\caption{\textbf{a} ABS spectrum of a SNS junction with two SC regions ($l_{SC} = 2000$ nm) separated by a normal region ($l = 80$ nm, $w = 1000$ nm) at $B = 2$ mT without disorder. The color curves denote analytically calculated ABS from Eq. \ref{ABS_energy_supp} with $\tau = 1$. \textbf{b} Supercurrent in the normal area of the junction calculated for the ABS whose energies are denoted by the color circles in \textbf{a}.}
\label{supp_figure_closed_system}
\end{figure}

For the numerical calculation of ABSs spectra we consider a Josephson junction treated as a finite system consisting of a normal scattering region and two long superconducting segments. The two superconducting regions have a length of $l_{SC} = 2000$~nm (much larger than the coherence length $\xi = 1091.16$ nm, calculated using the formula,  $\xi = \frac{\hbar v_{F}}{\Delta}$  where $v_{F} = \sqrt{{2\mu}/m^{*}}$ ), and they are separated by a normal region of length $l = 80$~nm. The width $w$ of the entire system is taken as 1000 nm.  The Hamiltonian remains the same as in equation \ref{eq:3} except for the tunneling barrier potentials (here we do not consider the top and bottom electrode). The anisotropic mass and Peierls phase factor (for magnetic vector potential) are introduced as described above. We diagonalize the Hamiltonian and plot the energy with respect to the phase difference $\phi$, and also the probability current in Fig. \ref{supp_figure_closed_system}. In the probability current, we observe that in the presence of the perpendicular field each ABSs is localized in a separated region in the junction. The different spatial position of the ABSs is reflected by their different phase shifts in the spectrum plotted in Fig. \ref{supp_figure_closed_system} (a).

\begin{thebibliography}{26}%
\makeatletter
\providecommand \@ifxundefined [1]{%
 \@ifx{#1\undefined}
}%
\providecommand \@ifnum [1]{%
 \ifnum #1\expandafter \@firstoftwo
 \else \expandafter \@secondoftwo
 \fi
}%
\providecommand \@ifx [1]{%
 \ifx #1\expandafter \@firstoftwo
 \else \expandafter \@secondoftwo
 \fi
}%
\providecommand \natexlab [1]{#1}%
\providecommand \enquote  [1]{``#1''}%
\providecommand \bibnamefont  [1]{#1}%
\providecommand \bibfnamefont [1]{#1}%
\providecommand \citenamefont [1]{#1}%
\providecommand \href@noop [0]{\@secondoftwo}%
\providecommand \href [0]{\begingroup \@sanitize@url \@href}%
\providecommand \@href[1]{\@@startlink{#1}\@@href}%
\providecommand \@@href[1]{\endgroup#1\@@endlink}%
\providecommand \@sanitize@url [0]{\catcode `\\12\catcode `\$12\catcode
  `\&12\catcode `\#12\catcode `\^12\catcode `\_12\catcode `\%12\relax}%
\providecommand \@@startlink[1]{}%
\providecommand \@@endlink[0]{}%
\providecommand \url  [0]{\begingroup\@sanitize@url \@url }%
\providecommand \@url [1]{\endgroup\@href {#1}{\urlprefix }}%
\providecommand \urlprefix  [0]{URL }%
\providecommand \Eprint [0]{\href }%
\providecommand \doibase [0]{https://doi.org/}%
\providecommand \selectlanguage [0]{\@gobble}%
\providecommand \bibinfo  [0]{\@secondoftwo}%
\providecommand \bibfield  [0]{\@secondoftwo}%
\providecommand \translation [1]{[#1]}%
\providecommand \BibitemOpen [0]{}%
\providecommand \bibitemStop [0]{}%
\providecommand \bibitemNoStop [0]{.\EOS\space}%
\providecommand \EOS [0]{\spacefactor3000\relax}%
\providecommand \BibitemShut  [1]{\csname bibitem#1\endcsname}%
\let\auto@bib@innerbib\@empty
\bibitem [{\citenamefont {Kulik}(1969)}]{kulik_1969}%
  \BibitemOpen
  \bibfield  {author} {\bibinfo {author} {\bibfnamefont {I.~O.}\ \bibnamefont
  {Kulik}},\ }\bibfield  {title} {\bibinfo {title} {Macroscopic quantization
  and the proximity effect in {S}{N}{S} junctions},\ }\href@noop {} {\bibfield
  {journal} {\bibinfo  {journal} {Sov. Phys. JETP}\ }\textbf {\bibinfo {volume}
  {30}},\ \bibinfo {pages} {944} (\bibinfo {year} {1969})}\BibitemShut
  {NoStop}%
\bibitem [{\citenamefont {Beenakker}(1991)}]{Beenakker_1991}%
  \BibitemOpen
  \bibfield  {author} {\bibinfo {author} {\bibfnamefont {C.~W.~J.}\
  \bibnamefont {Beenakker}},\ }\bibfield  {title} {\bibinfo {title} {Universal
  limit of critical-current fluctuations in mesoscopic {J}osephson junctions},\
  }\href {https://doi.org/10.1103/PhysRevLett.67.3836} {\bibfield  {journal}
  {\bibinfo  {journal} {Phys. Rev. Lett.}\ }\textbf {\bibinfo {volume} {67}},\
  \bibinfo {pages} {3836} (\bibinfo {year} {1991})}\BibitemShut {NoStop}%
\bibitem [{\citenamefont {Wendin}\ and\ \citenamefont
  {Shumeiko}(1996)}]{wendin_1996_josephson}%
  \BibitemOpen
  \bibfield  {author} {\bibinfo {author} {\bibfnamefont {G.}~\bibnamefont
  {Wendin}}\ and\ \bibinfo {author} {\bibfnamefont {V.~S.}\ \bibnamefont
  {Shumeiko}},\ }\bibfield  {title} {\bibinfo {title} {Josephson transport in
  complex mesoscopic structures},\ }\href
  {https://www.sciencedirect.com/science/article/pii/S0749603696901160}
  {\bibfield  {journal} {\bibinfo  {journal} {Superlattices Microstruct.}\
  }\textbf {\bibinfo {volume} {20}},\ \bibinfo {pages} {569} (\bibinfo {year}
  {1996})}\BibitemShut {NoStop}%
\bibitem [{\citenamefont {Su}\ \emph {et~al.}(2017)\citenamefont {Su},
  \citenamefont {Tacla}, \citenamefont {Hocevar}, \citenamefont {Car},
  \citenamefont {Plissard}, \citenamefont {Bakkers}, \citenamefont {Daley},
  \citenamefont {Pekker},\ and\ \citenamefont {Frolov}}]{su_2017}%
  \BibitemOpen
  \bibfield  {author} {\bibinfo {author} {\bibfnamefont {Z.}~\bibnamefont
  {Su}}, \bibinfo {author} {\bibfnamefont {A.~B.}\ \bibnamefont {Tacla}},
  \bibinfo {author} {\bibfnamefont {M.}~\bibnamefont {Hocevar}}, \bibinfo
  {author} {\bibfnamefont {D.}~\bibnamefont {Car}}, \bibinfo {author}
  {\bibfnamefont {S.~R.}\ \bibnamefont {Plissard}}, \bibinfo {author}
  {\bibfnamefont {E.~P.}\ \bibnamefont {Bakkers}}, \bibinfo {author}
  {\bibfnamefont {A.~J.}\ \bibnamefont {Daley}}, \bibinfo {author}
  {\bibfnamefont {D.}~\bibnamefont {Pekker}},\ and\ \bibinfo {author}
  {\bibfnamefont {S.~M.}\ \bibnamefont {Frolov}},\ }\bibfield  {title}
  {\bibinfo {title} {Andreev molecules in semiconductor nanowire double quantum
  dots},\ }\href {https://www.nature.com/articles/s41467-017-00665-7}
  {\bibfield  {journal} {\bibinfo  {journal} {Nat. Commun.}\ }\textbf {\bibinfo
  {volume} {8}},\ \bibinfo {pages} {1} (\bibinfo {year} {2017})}\BibitemShut
  {NoStop}%
\bibitem [{\citenamefont {Pillet}\ \emph {et~al.}(2019)\citenamefont {Pillet},
  \citenamefont {Benzoni}, \citenamefont {Griesmar}, \citenamefont {Smirr},\
  and\ \citenamefont {Girit}}]{pillet_2019}%
  \BibitemOpen
  \bibfield  {author} {\bibinfo {author} {\bibfnamefont {J.-D.}\ \bibnamefont
  {Pillet}}, \bibinfo {author} {\bibfnamefont {V.}~\bibnamefont {Benzoni}},
  \bibinfo {author} {\bibfnamefont {J.}~\bibnamefont {Griesmar}}, \bibinfo
  {author} {\bibfnamefont {J.-L.}\ \bibnamefont {Smirr}},\ and\ \bibinfo
  {author} {\bibfnamefont {{\c{C}}.~{\"{O}}.}\ \bibnamefont {Girit}},\
  }\bibfield  {title} {\bibinfo {title} {Nonlocal {J}osephson effect in
  {A}ndreev molecules},\ }\href
  {https://pubs.acs.org/doi/10.1021/acs.nanolett.9b02686} {\bibfield  {journal}
  {\bibinfo  {journal} {Nano Lett.}\ }\textbf {\bibinfo {volume} {19}},\
  \bibinfo {pages} {7138} (\bibinfo {year} {2019})}\BibitemShut {NoStop}%
\bibitem [{\citenamefont {K{\"u}rt{\"o}ssy}\ \emph {et~al.}(2021)\citenamefont
  {K{\"u}rt{\"o}ssy}, \citenamefont {Scher{\"u}bl}, \citenamefont
  {F{\"u}l{\"o}p}, \citenamefont {Luk{\'a}cs}, \citenamefont {Kanne},
  \citenamefont {Nyg{\aa}rd}, \citenamefont {Makk},\ and\ \citenamefont
  {Csonka}}]{kurtossy_2021}%
  \BibitemOpen
  \bibfield  {author} {\bibinfo {author} {\bibfnamefont {O.}~\bibnamefont
  {K{\"u}rt{\"o}ssy}}, \bibinfo {author} {\bibfnamefont {Z.}~\bibnamefont
  {Scher{\"u}bl}}, \bibinfo {author} {\bibfnamefont {G.}~\bibnamefont
  {F{\"u}l{\"o}p}}, \bibinfo {author} {\bibfnamefont {I.~E.}\ \bibnamefont
  {Luk{\'a}cs}}, \bibinfo {author} {\bibfnamefont {T.}~\bibnamefont {Kanne}},
  \bibinfo {author} {\bibfnamefont {J.}~\bibnamefont {Nyg{\aa}rd}}, \bibinfo
  {author} {\bibfnamefont {P.}~\bibnamefont {Makk}},\ and\ \bibinfo {author}
  {\bibfnamefont {S.}~\bibnamefont {Csonka}},\ }\bibfield  {title} {\bibinfo
  {title} {Andreev molecule in parallel {I}n{A}s nanowires},\ }\href
  {https://pubs.acs.org/doi/full/10.1021/acs.nanolett.1c01956} {\bibfield
  {journal} {\bibinfo  {journal} {Nano Lett.}\ }\textbf {\bibinfo {volume}
  {21}},\ \bibinfo {pages} {7929} (\bibinfo {year} {2021})}\BibitemShut
  {NoStop}%
\bibitem [{\citenamefont {Jünger}\ \emph {et~al.}(2021)\citenamefont
  {Jünger}, \citenamefont {Lehmann}, \citenamefont {Dick}, \citenamefont
  {Thelander}, \citenamefont {Schönenberger},\ and\ \citenamefont
  {Baumgartner}}]{Junger_2021}%
  \BibitemOpen
  \bibfield  {author} {\bibinfo {author} {\bibfnamefont {C.}~\bibnamefont
  {Jünger}}, \bibinfo {author} {\bibfnamefont {S.}~\bibnamefont {Lehmann}},
  \bibinfo {author} {\bibfnamefont {K.~A.}\ \bibnamefont {Dick}}, \bibinfo
  {author} {\bibfnamefont {C.}~\bibnamefont {Thelander}}, \bibinfo {author}
  {\bibfnamefont {C.}~\bibnamefont {Schönenberger}},\ and\ \bibinfo {author}
  {\bibfnamefont {A.}~\bibnamefont {Baumgartner}},\ }\href
  {https://doi.org/10.48550/ARXIV.2111.00651} {\bibinfo {title} {Intermediate
  states in andreev bound state fusion}} (\bibinfo {year} {2021}),\ \bibinfo
  {note} {(accessed July 13, 2022)}\BibitemShut {NoStop}%
\bibitem [{\citenamefont {Hays}\ \emph {et~al.}(2018)\citenamefont {Hays},
  \citenamefont {de~Lange}, \citenamefont {Serniak}, \citenamefont {van
  Woerkom}, \citenamefont {Bouman}, \citenamefont {Krogstrup}, \citenamefont
  {Nyg\aa{}rd}, \citenamefont {Geresdi},\ and\ \citenamefont
  {Devoret}}]{Hays_2018}%
  \BibitemOpen
  \bibfield  {author} {\bibinfo {author} {\bibfnamefont {M.}~\bibnamefont
  {Hays}}, \bibinfo {author} {\bibfnamefont {G.}~\bibnamefont {de~Lange}},
  \bibinfo {author} {\bibfnamefont {K.}~\bibnamefont {Serniak}}, \bibinfo
  {author} {\bibfnamefont {D.~J.}\ \bibnamefont {van Woerkom}}, \bibinfo
  {author} {\bibfnamefont {D.}~\bibnamefont {Bouman}}, \bibinfo {author}
  {\bibfnamefont {P.}~\bibnamefont {Krogstrup}}, \bibinfo {author}
  {\bibfnamefont {J.}~\bibnamefont {Nyg\aa{}rd}}, \bibinfo {author}
  {\bibfnamefont {A.}~\bibnamefont {Geresdi}},\ and\ \bibinfo {author}
  {\bibfnamefont {M.~H.}\ \bibnamefont {Devoret}},\ }\bibfield  {title}
  {\bibinfo {title} {Direct microwave measurement of {A}ndreev-bound-state
  dynamics in a semiconductor-nanowire {J}osephson junction},\ }\href
  {https://doi.org/10.1103/PhysRevLett.121.047001} {\bibfield  {journal}
  {\bibinfo  {journal} {Phys. Rev. Lett.}\ }\textbf {\bibinfo {volume} {121}},\
  \bibinfo {pages} {047001} (\bibinfo {year} {2018})}\BibitemShut {NoStop}%
\bibitem [{\citenamefont {Hays}\ \emph {et~al.}(2021)\citenamefont {Hays},
  \citenamefont {Fatemi}, \citenamefont {Bouman}, \citenamefont {Cerrillo},
  \citenamefont {Diamond}, \citenamefont {Serniak}, \citenamefont {Connolly},
  \citenamefont {Krogstrup}, \citenamefont {Nygård}, \citenamefont {Yeyati},
  \citenamefont {Geresdi},\ and\ \citenamefont {Devoret}}]{Hays_2021}%
  \BibitemOpen
  \bibfield  {author} {\bibinfo {author} {\bibfnamefont {M.}~\bibnamefont
  {Hays}}, \bibinfo {author} {\bibfnamefont {V.}~\bibnamefont {Fatemi}},
  \bibinfo {author} {\bibfnamefont {D.}~\bibnamefont {Bouman}}, \bibinfo
  {author} {\bibfnamefont {J.}~\bibnamefont {Cerrillo}}, \bibinfo {author}
  {\bibfnamefont {S.}~\bibnamefont {Diamond}}, \bibinfo {author} {\bibfnamefont
  {K.}~\bibnamefont {Serniak}}, \bibinfo {author} {\bibfnamefont
  {T.}~\bibnamefont {Connolly}}, \bibinfo {author} {\bibfnamefont
  {P.}~\bibnamefont {Krogstrup}}, \bibinfo {author} {\bibfnamefont
  {J.}~\bibnamefont {Nygård}}, \bibinfo {author} {\bibfnamefont {A.~L.}\
  \bibnamefont {Yeyati}}, \bibinfo {author} {\bibfnamefont {A.}~\bibnamefont
  {Geresdi}},\ and\ \bibinfo {author} {\bibfnamefont {M.~H.}\ \bibnamefont
  {Devoret}},\ }\bibfield  {title} {\bibinfo {title} {Coherent manipulation of
  an {A}ndreev spin qubit},\ }\href {https://doi.org/10.1126/science.abf0345}
  {\bibfield  {journal} {\bibinfo  {journal} {Science}\ }\textbf {\bibinfo
  {volume} {373}},\ \bibinfo {pages} {430} (\bibinfo {year}
  {2021})}\BibitemShut {NoStop}%
\bibitem [{\citenamefont {Tinkham}(1996)}]{Tinkham_1996}%
  \BibitemOpen
  \bibfield  {author} {\bibinfo {author} {\bibfnamefont {M.}~\bibnamefont
  {Tinkham}},\ }\href@noop {} {\emph {\bibinfo {title} {Introduction to
  Superconductivity}}},\ \bibinfo {edition} {2nd}\ ed.\ (\bibinfo  {publisher}
  {Dover Publications},\ \bibinfo {year} {1996})\ pp.\ \bibinfo {pages}
  {215--220}\BibitemShut {NoStop}%
\bibitem [{\citenamefont {Cuevas}\ and\ \citenamefont
  {Bergeret}(2007)}]{Cuevas_2007_Magnetic}%
  \BibitemOpen
  \bibfield  {author} {\bibinfo {author} {\bibfnamefont {J.~C.}\ \bibnamefont
  {Cuevas}}\ and\ \bibinfo {author} {\bibfnamefont {F.~S.}\ \bibnamefont
  {Bergeret}},\ }\bibfield  {title} {\bibinfo {title} {Magnetic interference
  patterns and vortices in diffusive {SNS} junctions},\ }\href
  {https://doi.org/10.1103/PhysRevLett.99.217002} {\bibfield  {journal}
  {\bibinfo  {journal} {Phys. Rev. Lett.}\ }\textbf {\bibinfo {volume} {99}},\
  \bibinfo {pages} {217002} (\bibinfo {year} {2007})}\BibitemShut {NoStop}%
\bibitem [{\citenamefont {Kaperek}\ \emph {et~al.}(2022)\citenamefont
  {Kaperek}, \citenamefont {Heun}, \citenamefont {Carrega}, \citenamefont
  {W\'ojcik},\ and\ \citenamefont {Nowak}}]{Kaperek_2022}%
  \BibitemOpen
  \bibfield  {author} {\bibinfo {author} {\bibfnamefont {K.}~\bibnamefont
  {Kaperek}}, \bibinfo {author} {\bibfnamefont {S.}~\bibnamefont {Heun}},
  \bibinfo {author} {\bibfnamefont {M.}~\bibnamefont {Carrega}}, \bibinfo
  {author} {\bibfnamefont {P.}~\bibnamefont {W\'ojcik}},\ and\ \bibinfo
  {author} {\bibfnamefont {M.~P.}\ \bibnamefont {Nowak}},\ }\bibfield  {title}
  {\bibinfo {title} {Theory of scanning gate microscopy imaging of the
  supercurrent distribution in a planar josephson junction},\ }\href
  {https://doi.org/10.1103/PhysRevB.106.035432} {\bibfield  {journal} {\bibinfo
   {journal} {Phys. Rev. B}\ }\textbf {\bibinfo {volume} {106}},\ \bibinfo
  {pages} {035432} (\bibinfo {year} {2022})}\BibitemShut {NoStop}%
\bibitem [{\citenamefont {Hell}\ \emph {et~al.}(2017)\citenamefont {Hell},
  \citenamefont {Leijnse},\ and\ \citenamefont {Flensberg}}]{Hell_2017}%
  \BibitemOpen
  \bibfield  {author} {\bibinfo {author} {\bibfnamefont {M.}~\bibnamefont
  {Hell}}, \bibinfo {author} {\bibfnamefont {M.}~\bibnamefont {Leijnse}},\ and\
  \bibinfo {author} {\bibfnamefont {K.}~\bibnamefont {Flensberg}},\ }\bibfield
  {title} {\bibinfo {title} {Two-dimensional platform for networks of
  {M}ajorana bound states},\ }\href
  {https://doi.org/10.1103/PhysRevLett.118.107701} {\bibfield  {journal}
  {\bibinfo  {journal} {Phys. Rev. Lett.}\ }\textbf {\bibinfo {volume} {118}},\
  \bibinfo {pages} {107701} (\bibinfo {year} {2017})}\BibitemShut {NoStop}%
\bibitem [{\citenamefont {Pientka}\ \emph {et~al.}(2017)\citenamefont
  {Pientka}, \citenamefont {Keselman}, \citenamefont {Berg}, \citenamefont
  {Yacoby}, \citenamefont {Stern},\ and\ \citenamefont
  {Halperin}}]{Pientka_2017}%
  \BibitemOpen
  \bibfield  {author} {\bibinfo {author} {\bibfnamefont {F.}~\bibnamefont
  {Pientka}}, \bibinfo {author} {\bibfnamefont {A.}~\bibnamefont {Keselman}},
  \bibinfo {author} {\bibfnamefont {E.}~\bibnamefont {Berg}}, \bibinfo {author}
  {\bibfnamefont {A.}~\bibnamefont {Yacoby}}, \bibinfo {author} {\bibfnamefont
  {A.}~\bibnamefont {Stern}},\ and\ \bibinfo {author} {\bibfnamefont {B.~I.}\
  \bibnamefont {Halperin}},\ }\bibfield  {title} {\bibinfo {title} {Topological
  superconductivity in a planar {J}osephson junction},\ }\href
  {https://doi.org/10.1103/PhysRevX.7.021032} {\bibfield  {journal} {\bibinfo
  {journal} {Phys. Rev. X}\ }\textbf {\bibinfo {volume} {7}},\ \bibinfo {pages}
  {021032} (\bibinfo {year} {2017})}\BibitemShut {NoStop}%
\bibitem [{\citenamefont {Ren}\ \emph {et~al.}(2019)\citenamefont {Ren},
  \citenamefont {Pientka}, \citenamefont {Hart}, \citenamefont {Pierce},
  \citenamefont {Kosowsky}, \citenamefont {Lunczer}, \citenamefont {Schlereth},
  \citenamefont {Scharf}, \citenamefont {Hankiewicz}, \citenamefont
  {Molenkamp}, \citenamefont {Halperin},\ and\ \citenamefont
  {Yacoby}}]{Ren_2019}%
  \BibitemOpen
  \bibfield  {author} {\bibinfo {author} {\bibfnamefont {H.}~\bibnamefont
  {Ren}}, \bibinfo {author} {\bibfnamefont {F.}~\bibnamefont {Pientka}},
  \bibinfo {author} {\bibfnamefont {S.}~\bibnamefont {Hart}}, \bibinfo {author}
  {\bibfnamefont {A.~T.}\ \bibnamefont {Pierce}}, \bibinfo {author}
  {\bibfnamefont {M.}~\bibnamefont {Kosowsky}}, \bibinfo {author}
  {\bibfnamefont {L.}~\bibnamefont {Lunczer}}, \bibinfo {author} {\bibfnamefont
  {R.}~\bibnamefont {Schlereth}}, \bibinfo {author} {\bibfnamefont
  {B.}~\bibnamefont {Scharf}}, \bibinfo {author} {\bibfnamefont {E.~M.}\
  \bibnamefont {Hankiewicz}}, \bibinfo {author} {\bibfnamefont {L.~W.}\
  \bibnamefont {Molenkamp}}, \bibinfo {author} {\bibfnamefont {B.~I.}\
  \bibnamefont {Halperin}},\ and\ \bibinfo {author} {\bibfnamefont
  {A.}~\bibnamefont {Yacoby}},\ }\bibfield  {title} {\bibinfo {title}
  {Topological superconductivity in a phase-controlled {J}osephson junction},\
  }\href {https://doi.org/10.1038/s41586-019-1148-9} {\bibfield  {journal}
  {\bibinfo  {journal} {Nature}\ }\textbf {\bibinfo {volume} {569}},\ \bibinfo
  {pages} {93} (\bibinfo {year} {2019})}\BibitemShut {NoStop}%
\bibitem [{\citenamefont {Fornieri}\ \emph {et~al.}(2019)\citenamefont
  {Fornieri}, \citenamefont {Whiticar}, \citenamefont {Setiawan}, \citenamefont
  {Portol{\'e}s}, \citenamefont {Drachmann}, \citenamefont {Keselman},
  \citenamefont {Gronin}, \citenamefont {Thomas}, \citenamefont {Wang},
  \citenamefont {Kallaher}, \citenamefont {Gardner}, \citenamefont {Berg},
  \citenamefont {Manfra}, \citenamefont {Stern}, \citenamefont {Marcus},\ and\
  \citenamefont {Nichele}}]{Fornieri_2019}%
  \BibitemOpen
  \bibfield  {author} {\bibinfo {author} {\bibfnamefont {A.}~\bibnamefont
  {Fornieri}}, \bibinfo {author} {\bibfnamefont {A.~M.}\ \bibnamefont
  {Whiticar}}, \bibinfo {author} {\bibfnamefont {F.}~\bibnamefont {Setiawan}},
  \bibinfo {author} {\bibfnamefont {E.}~\bibnamefont {Portol{\'e}s}}, \bibinfo
  {author} {\bibfnamefont {A.~C.~C.}\ \bibnamefont {Drachmann}}, \bibinfo
  {author} {\bibfnamefont {A.}~\bibnamefont {Keselman}}, \bibinfo {author}
  {\bibfnamefont {S.}~\bibnamefont {Gronin}}, \bibinfo {author} {\bibfnamefont
  {C.}~\bibnamefont {Thomas}}, \bibinfo {author} {\bibfnamefont
  {T.}~\bibnamefont {Wang}}, \bibinfo {author} {\bibfnamefont {R.}~\bibnamefont
  {Kallaher}}, \bibinfo {author} {\bibfnamefont {G.~C.}\ \bibnamefont
  {Gardner}}, \bibinfo {author} {\bibfnamefont {E.}~\bibnamefont {Berg}},
  \bibinfo {author} {\bibfnamefont {M.~J.}\ \bibnamefont {Manfra}}, \bibinfo
  {author} {\bibfnamefont {A.}~\bibnamefont {Stern}}, \bibinfo {author}
  {\bibfnamefont {C.~M.}\ \bibnamefont {Marcus}},\ and\ \bibinfo {author}
  {\bibfnamefont {F.}~\bibnamefont {Nichele}},\ }\bibfield  {title} {\bibinfo
  {title} {Evidence of topological superconductivity in planar {J}osephson
  junctions},\ }\href {https://doi.org/10.1038/s41586-019-1068-8} {\bibfield
  {journal} {\bibinfo  {journal} {Nature}\ }\textbf {\bibinfo {volume} {569}},\
  \bibinfo {pages} {89} (\bibinfo {year} {2019})}\BibitemShut {NoStop}%
\bibitem [{\citenamefont {Stern}\ and\ \citenamefont
  {Berg}(2019)}]{Stern_2019}%
  \BibitemOpen
  \bibfield  {author} {\bibinfo {author} {\bibfnamefont {A.}~\bibnamefont
  {Stern}}\ and\ \bibinfo {author} {\bibfnamefont {E.}~\bibnamefont {Berg}},\
  }\bibfield  {title} {\bibinfo {title} {Fractional {J}osephson vortices and
  braiding of {M}ajorana zero modes in planar superconductor-semiconductor
  heterostructures},\ }\href {https://doi.org/10.1103/PhysRevLett.122.107701}
  {\bibfield  {journal} {\bibinfo  {journal} {Phys. Rev. Lett.}\ }\textbf
  {\bibinfo {volume} {122}},\ \bibinfo {pages} {107701} (\bibinfo {year}
  {2019})}\BibitemShut {NoStop}%
\bibitem [{\citenamefont {le~Sueur}\ \emph {et~al.}(2008)\citenamefont
  {le~Sueur}, \citenamefont {Joyez}, \citenamefont {Pothier}, \citenamefont
  {Urbina},\ and\ \citenamefont {Esteve}}]{leSueur_2008}%
  \BibitemOpen
  \bibfield  {author} {\bibinfo {author} {\bibfnamefont {H.}~\bibnamefont
  {le~Sueur}}, \bibinfo {author} {\bibfnamefont {P.}~\bibnamefont {Joyez}},
  \bibinfo {author} {\bibfnamefont {H.}~\bibnamefont {Pothier}}, \bibinfo
  {author} {\bibfnamefont {C.}~\bibnamefont {Urbina}},\ and\ \bibinfo {author}
  {\bibfnamefont {D.}~\bibnamefont {Esteve}},\ }\bibfield  {title} {\bibinfo
  {title} {Phase controlled superconducting proximity effect probed by
  tunneling spectroscopy},\ }\href
  {https://doi.org/10.1103/PhysRevLett.100.197002} {\bibfield  {journal}
  {\bibinfo  {journal} {Phys. Rev. Lett.}\ }\textbf {\bibinfo {volume} {100}},\
  \bibinfo {pages} {197002} (\bibinfo {year} {2008})}\BibitemShut {NoStop}%
\bibitem [{\citenamefont {Roditchev}\ \emph {et~al.}(2015)\citenamefont
  {Roditchev}, \citenamefont {Brun}, \citenamefont {Serrier-Garcia},
  \citenamefont {Cuevas}, \citenamefont {Bessa}, \citenamefont
  {Milo{\v{s}}evi{\'{c}}}, \citenamefont {Debontridder}, \citenamefont
  {Stolyarov},\ and\ \citenamefont {Cren}}]{Roditchev_2015}%
  \BibitemOpen
  \bibfield  {author} {\bibinfo {author} {\bibfnamefont {D.}~\bibnamefont
  {Roditchev}}, \bibinfo {author} {\bibfnamefont {C.}~\bibnamefont {Brun}},
  \bibinfo {author} {\bibfnamefont {L.}~\bibnamefont {Serrier-Garcia}},
  \bibinfo {author} {\bibfnamefont {J.~C.}\ \bibnamefont {Cuevas}}, \bibinfo
  {author} {\bibfnamefont {V.~H.~L.}\ \bibnamefont {Bessa}}, \bibinfo {author}
  {\bibfnamefont {M.~V.}\ \bibnamefont {Milo{\v{s}}evi{\'{c}}}}, \bibinfo
  {author} {\bibfnamefont {F.}~\bibnamefont {Debontridder}}, \bibinfo {author}
  {\bibfnamefont {V.}~\bibnamefont {Stolyarov}},\ and\ \bibinfo {author}
  {\bibfnamefont {T.}~\bibnamefont {Cren}},\ }\bibfield  {title} {\bibinfo
  {title} {Direct observation of {J}osephson vortex cores},\ }\href
  {https://doi.org/10.1038/nphys3240} {\bibfield  {journal} {\bibinfo
  {journal} {Nat. Phys.}\ }\textbf {\bibinfo {volume} {11}},\ \bibinfo {pages}
  {332} (\bibinfo {year} {2015})}\BibitemShut {NoStop}%
\bibitem [{\citenamefont {Banerjee}\ \emph
  {et~al.}(2022{\natexlab{a}})\citenamefont {Banerjee}, \citenamefont {Lesser},
  \citenamefont {Rahman}, \citenamefont {Wang}, \citenamefont {Li},
  \citenamefont {Kringhøj}, \citenamefont {Whiticar}, \citenamefont
  {Drachmann}, \citenamefont {Thomas}, \citenamefont {Wang}, \citenamefont
  {Manfra}, \citenamefont {Berg}, \citenamefont {Oreg}, \citenamefont {Stern},\
  and\ \citenamefont {Marcus}}]{Banerjee_2022_topological_phase}%
  \BibitemOpen
  \bibfield  {author} {\bibinfo {author} {\bibfnamefont {A.}~\bibnamefont
  {Banerjee}}, \bibinfo {author} {\bibfnamefont {O.}~\bibnamefont {Lesser}},
  \bibinfo {author} {\bibfnamefont {M.~A.}\ \bibnamefont {Rahman}}, \bibinfo
  {author} {\bibfnamefont {H.~R.}\ \bibnamefont {Wang}}, \bibinfo {author}
  {\bibfnamefont {M.~R.}\ \bibnamefont {Li}}, \bibinfo {author} {\bibfnamefont
  {A.}~\bibnamefont {Kringhøj}}, \bibinfo {author} {\bibfnamefont {A.~M.}\
  \bibnamefont {Whiticar}}, \bibinfo {author} {\bibfnamefont {A.~C.~C.}\
  \bibnamefont {Drachmann}}, \bibinfo {author} {\bibfnamefont {C.}~\bibnamefont
  {Thomas}}, \bibinfo {author} {\bibfnamefont {T.}~\bibnamefont {Wang}},
  \bibinfo {author} {\bibfnamefont {M.~J.}\ \bibnamefont {Manfra}}, \bibinfo
  {author} {\bibfnamefont {E.}~\bibnamefont {Berg}}, \bibinfo {author}
  {\bibfnamefont {Y.}~\bibnamefont {Oreg}}, \bibinfo {author} {\bibfnamefont
  {A.}~\bibnamefont {Stern}},\ and\ \bibinfo {author} {\bibfnamefont {C.~M.}\
  \bibnamefont {Marcus}},\ }\href {https://doi.org/10.48550/ARXIV.2201.03453}
  {\bibinfo {title} {Signatures of a topological phase transition in a planar
  josephson junction}} (\bibinfo {year} {2022}{\natexlab{a}}),\ \bibinfo {note}
  {(accessed July 13, 2022)}\BibitemShut {NoStop}%
\bibitem [{\citenamefont {Banerjee}\ \emph
  {et~al.}(2022{\natexlab{b}})\citenamefont {Banerjee}, \citenamefont {Lesser},
  \citenamefont {Rahman}, \citenamefont {Thomas}, \citenamefont {Wang},
  \citenamefont {Manfra}, \citenamefont {Berg}, \citenamefont {Oreg},
  \citenamefont {Stern},\ and\ \citenamefont
  {Marcus}}]{Banerjee_2022_Local_Nonlocal}%
  \BibitemOpen
  \bibfield  {author} {\bibinfo {author} {\bibfnamefont {A.}~\bibnamefont
  {Banerjee}}, \bibinfo {author} {\bibfnamefont {O.}~\bibnamefont {Lesser}},
  \bibinfo {author} {\bibfnamefont {M.~A.}\ \bibnamefont {Rahman}}, \bibinfo
  {author} {\bibfnamefont {C.}~\bibnamefont {Thomas}}, \bibinfo {author}
  {\bibfnamefont {T.}~\bibnamefont {Wang}}, \bibinfo {author} {\bibfnamefont
  {M.~J.}\ \bibnamefont {Manfra}}, \bibinfo {author} {\bibfnamefont
  {E.}~\bibnamefont {Berg}}, \bibinfo {author} {\bibfnamefont {Y.}~\bibnamefont
  {Oreg}}, \bibinfo {author} {\bibfnamefont {A.}~\bibnamefont {Stern}},\ and\
  \bibinfo {author} {\bibfnamefont {C.~M.}\ \bibnamefont {Marcus}},\ }\href
  {https://doi.org/10.48550/ARXIV.2205.09419} {\bibinfo {title} {Local and
  nonlocal transport spectroscopy in planar josephson junctions}} (\bibinfo
  {year} {2022}{\natexlab{b}}),\ \bibinfo {note} {(accessed July 13,
  2022)}\BibitemShut {NoStop}%
\bibitem [{\citenamefont {Moehle}\ \emph {et~al.}(2021)\citenamefont {Moehle},
  \citenamefont {Ke}, \citenamefont {Wang}, \citenamefont {Thomas},
  \citenamefont {Xiao}, \citenamefont {Karwal}, \citenamefont {Lodari},
  \citenamefont {van~de Kerkhof}, \citenamefont {Termaat}, \citenamefont
  {Gardner}, \citenamefont {Scappucci}, \citenamefont {Manfra},\ and\
  \citenamefont {Goswami}}]{Moehle_2021}%
  \BibitemOpen
  \bibfield  {author} {\bibinfo {author} {\bibfnamefont {C.~M.}\ \bibnamefont
  {Moehle}}, \bibinfo {author} {\bibfnamefont {C.~T.}\ \bibnamefont {Ke}},
  \bibinfo {author} {\bibfnamefont {Q.}~\bibnamefont {Wang}}, \bibinfo {author}
  {\bibfnamefont {C.}~\bibnamefont {Thomas}}, \bibinfo {author} {\bibfnamefont
  {D.}~\bibnamefont {Xiao}}, \bibinfo {author} {\bibfnamefont {S.}~\bibnamefont
  {Karwal}}, \bibinfo {author} {\bibfnamefont {M.}~\bibnamefont {Lodari}},
  \bibinfo {author} {\bibfnamefont {V.}~\bibnamefont {van~de Kerkhof}},
  \bibinfo {author} {\bibfnamefont {R.}~\bibnamefont {Termaat}}, \bibinfo
  {author} {\bibfnamefont {G.~C.}\ \bibnamefont {Gardner}}, \bibinfo {author}
  {\bibfnamefont {G.}~\bibnamefont {Scappucci}}, \bibinfo {author}
  {\bibfnamefont {M.~J.}\ \bibnamefont {Manfra}},\ and\ \bibinfo {author}
  {\bibfnamefont {S.}~\bibnamefont {Goswami}},\ }\bibfield  {title} {\bibinfo
  {title} {In{S}b{A}s two-dimensional electron gases as a platform for
  topological superconductivity},\ }\href
  {https://doi.org/10.1021/acs.nanolett.1c03520} {\bibfield  {journal}
  {\bibinfo  {journal} {Nano Lett.}\ }\textbf {\bibinfo {volume} {21}},\
  \bibinfo {pages} {9990} (\bibinfo {year} {2021})}\BibitemShut {NoStop}%
\bibitem [{\citenamefont {Newrock}\ \emph {et~al.}(2000)\citenamefont
  {Newrock}, \citenamefont {Lobb}, \citenamefont {Geigenmuller},\ and\
  \citenamefont {Octavio}}]{Newrock}%
  \BibitemOpen
  \bibfield  {author} {\bibinfo {author} {\bibfnamefont {R.}~\bibnamefont
  {Newrock}}, \bibinfo {author} {\bibfnamefont {C.}~\bibnamefont {Lobb}},
  \bibinfo {author} {\bibfnamefont {U.}~\bibnamefont {Geigenmuller}},\ and\
  \bibinfo {author} {\bibfnamefont {M.}~\bibnamefont {Octavio}},\ }\bibfield
  {title} {\bibinfo {title} {The two-dimensional physics of {J}osephson
  junction arrays},\ }\href@noop {} {\bibfield  {journal} {\bibinfo  {journal}
  {Solid State Physics}\ }\textbf {\bibinfo {volume} {54}},\ \bibinfo {pages}
  {263} (\bibinfo {year} {2000})}\BibitemShut {NoStop}%
\bibitem [{\citenamefont {Suominen}\ \emph {et~al.}(2017)\citenamefont
  {Suominen}, \citenamefont {Danon}, \citenamefont {Kjaergaard}, \citenamefont
  {Flensberg}, \citenamefont {Shabani}, \citenamefont {Palmstr\o{}m},
  \citenamefont {Nichele},\ and\ \citenamefont {Marcus}}]{Suominen_2017}%
  \BibitemOpen
  \bibfield  {author} {\bibinfo {author} {\bibfnamefont {H.~J.}\ \bibnamefont
  {Suominen}}, \bibinfo {author} {\bibfnamefont {J.}~\bibnamefont {Danon}},
  \bibinfo {author} {\bibfnamefont {M.}~\bibnamefont {Kjaergaard}}, \bibinfo
  {author} {\bibfnamefont {K.}~\bibnamefont {Flensberg}}, \bibinfo {author}
  {\bibfnamefont {J.}~\bibnamefont {Shabani}}, \bibinfo {author} {\bibfnamefont
  {C.~J.}\ \bibnamefont {Palmstr\o{}m}}, \bibinfo {author} {\bibfnamefont
  {F.}~\bibnamefont {Nichele}},\ and\ \bibinfo {author} {\bibfnamefont {C.~M.}\
  \bibnamefont {Marcus}},\ }\bibfield  {title} {\bibinfo {title} {Anomalous
  {F}raunhofer interference in epitaxial superconductor-semiconductor
  {J}osephson junctions},\ }\href {https://doi.org/10.1103/PhysRevB.95.035307}
  {\bibfield  {journal} {\bibinfo  {journal} {Phys. Rev. B}\ }\textbf {\bibinfo
  {volume} {95}},\ \bibinfo {pages} {035307} (\bibinfo {year}
  {2017})}\BibitemShut {NoStop}%
\bibitem [{\citenamefont {Nichele}\ \emph {et~al.}(2020)\citenamefont
  {Nichele}, \citenamefont {Portol\'es}, \citenamefont {Fornieri},
  \citenamefont {Whiticar}, \citenamefont {Drachmann}, \citenamefont {Gronin},
  \citenamefont {Wang}, \citenamefont {Gardner}, \citenamefont {Thomas},
  \citenamefont {Hatke}, \citenamefont {Manfra},\ and\ \citenamefont
  {Marcus}}]{Nichele_2020}%
  \BibitemOpen
  \bibfield  {author} {\bibinfo {author} {\bibfnamefont {F.}~\bibnamefont
  {Nichele}}, \bibinfo {author} {\bibfnamefont {E.}~\bibnamefont {Portol\'es}},
  \bibinfo {author} {\bibfnamefont {A.}~\bibnamefont {Fornieri}}, \bibinfo
  {author} {\bibfnamefont {A.~M.}\ \bibnamefont {Whiticar}}, \bibinfo {author}
  {\bibfnamefont {A.~C.~C.}\ \bibnamefont {Drachmann}}, \bibinfo {author}
  {\bibfnamefont {S.}~\bibnamefont {Gronin}}, \bibinfo {author} {\bibfnamefont
  {T.}~\bibnamefont {Wang}}, \bibinfo {author} {\bibfnamefont {G.~C.}\
  \bibnamefont {Gardner}}, \bibinfo {author} {\bibfnamefont {C.}~\bibnamefont
  {Thomas}}, \bibinfo {author} {\bibfnamefont {A.~T.}\ \bibnamefont {Hatke}},
  \bibinfo {author} {\bibfnamefont {M.~J.}\ \bibnamefont {Manfra}},\ and\
  \bibinfo {author} {\bibfnamefont {C.~M.}\ \bibnamefont {Marcus}},\ }\bibfield
   {title} {\bibinfo {title} {Relating {A}ndreev bound states and supercurrents
  in hybrid {J}osephson junctions},\ }\href
  {https://doi.org/10.1103/PhysRevLett.124.226801} {\bibfield  {journal}
  {\bibinfo  {journal} {Phys. Rev. Lett.}\ }\textbf {\bibinfo {volume} {124}},\
  \bibinfo {pages} {226801} (\bibinfo {year} {2020})}\BibitemShut {NoStop}%
\bibitem [{\citenamefont {Banerjee}\ \emph
  {et~al.}(2022{\natexlab{c}})\citenamefont {Banerjee}, \citenamefont {Geier},
  \citenamefont {Rahman}, \citenamefont {Sanchez}, \citenamefont {Thomas},
  \citenamefont {Wang}, \citenamefont {Manfra}, \citenamefont {Flensberg},\
  and\ \citenamefont {Marcus}}]{Banerjee_2022_phase_texture}%
  \BibitemOpen
  \bibfield  {author} {\bibinfo {author} {\bibfnamefont {A.}~\bibnamefont
  {Banerjee}}, \bibinfo {author} {\bibfnamefont {M.}~\bibnamefont {Geier}},
  \bibinfo {author} {\bibfnamefont {M.~A.}\ \bibnamefont {Rahman}}, \bibinfo
  {author} {\bibfnamefont {D.~S.}\ \bibnamefont {Sanchez}}, \bibinfo {author}
  {\bibfnamefont {C.}~\bibnamefont {Thomas}}, \bibinfo {author} {\bibfnamefont
  {T.}~\bibnamefont {Wang}}, \bibinfo {author} {\bibfnamefont {M.~J.}\
  \bibnamefont {Manfra}}, \bibinfo {author} {\bibfnamefont {K.}~\bibnamefont
  {Flensberg}},\ and\ \bibinfo {author} {\bibfnamefont {C.~M.}\ \bibnamefont
  {Marcus}},\ }\href {https://doi.org/10.48550/ARXIV.2205.15690} {\bibinfo
  {title} {Control of andreev bound states using superconducting phase
  texture}} (\bibinfo {year} {2022}{\natexlab{c}}),\ \bibinfo {note} {(accessed
  July 13, 2022)}\BibitemShut {NoStop}%
\end{thebibliography}

\clearpage

\begin{thebibliography}{12}%
\makeatletter
\providecommand \@ifxundefined [1]{%
 \@ifx{#1\undefined}
}%
\providecommand \@ifnum [1]{%
 \ifnum #1\expandafter \@firstoftwo
 \else \expandafter \@secondoftwo
 \fi
}%
\providecommand \@ifx [1]{%
 \ifx #1\expandafter \@firstoftwo
 \else \expandafter \@secondoftwo
 \fi
}%
\providecommand \natexlab [1]{#1}%
\providecommand \enquote  [1]{``#1''}%
\providecommand \bibnamefont  [1]{#1}%
\providecommand \bibfnamefont [1]{#1}%
\providecommand \citenamefont [1]{#1}%
\providecommand \href@noop [0]{\@secondoftwo}%
\providecommand \href [0]{\begingroup \@sanitize@url \@href}%
\providecommand \@href[1]{\@@startlink{#1}\@@href}%
\providecommand \@@href[1]{\endgroup#1\@@endlink}%
\providecommand \@sanitize@url [0]{\catcode `\\12\catcode `\$12\catcode
  `\&12\catcode `\#12\catcode `\^12\catcode `\_12\catcode `\%12\relax}%
\providecommand \@@startlink[1]{}%
\providecommand \@@endlink[0]{}%
\providecommand \url  [0]{\begingroup\@sanitize@url \@url }%
\providecommand \@url [1]{\endgroup\@href {#1}{\urlprefix }}%
\providecommand \urlprefix  [0]{URL }%
\providecommand \Eprint [0]{\href }%
\providecommand \doibase [0]{https://doi.org/}%
\providecommand \selectlanguage [0]{\@gobble}%
\providecommand \bibinfo  [0]{\@secondoftwo}%
\providecommand \bibfield  [0]{\@secondoftwo}%
\providecommand \translation [1]{[#1]}%
\providecommand \BibitemOpen [0]{}%
\providecommand \bibitemStop [0]{}%
\providecommand \bibitemNoStop [0]{.\EOS\space}%
\providecommand \EOS [0]{\spacefactor3000\relax}%
\providecommand \BibitemShut  [1]{\csname bibitem#1\endcsname}%
\let\auto@bib@innerbib\@empty
\bibitem [{\citenamefont {Laeven}\ \emph {et~al.}(2020)\citenamefont {Laeven},
  \citenamefont {Nijholt}, \citenamefont {Wimmer},\ and\ \citenamefont
  {Akhmerov}}]{Laeven_2020}%
  \BibitemOpen
  \bibfield  {author} {\bibinfo {author} {\bibfnamefont {T.}~\bibnamefont
  {Laeven}}, \bibinfo {author} {\bibfnamefont {B.}~\bibnamefont {Nijholt}},
  \bibinfo {author} {\bibfnamefont {M.}~\bibnamefont {Wimmer}},\ and\ \bibinfo
  {author} {\bibfnamefont {A.~R.}\ \bibnamefont {Akhmerov}},\ }\bibfield
  {title} {\bibinfo {title} {Enhanced proximity effect in zigzag-shaped
  majorana josephson junctions},\ }\href
  {https://doi.org/10.1103/PhysRevLett.125.086802} {\bibfield  {journal}
  {\bibinfo  {journal} {Phys. Rev. Lett.}\ }\textbf {\bibinfo {volume} {125}},\
  \bibinfo {pages} {086802} (\bibinfo {year} {2020})}\BibitemShut {NoStop}%
\bibitem [{\citenamefont {Braginski}\ and\ \citenamefont
  {Clarke}(2004)}]{Clarke_2004}%
  \BibitemOpen
  \bibfield  {author} {\bibinfo {author} {\bibfnamefont {A.~I.}\ \bibnamefont
  {Braginski}}\ and\ \bibinfo {author} {\bibfnamefont {J.}~\bibnamefont
  {Clarke}},\ }\href@noop {} {\emph {\bibinfo {title} {The SQUID Handbook}}}\
  (\bibinfo  {publisher} {John Wiley and Sons, Ltd},\ \bibinfo {year}
  {2004})\BibitemShut {NoStop}%
\bibitem [{\citenamefont {Suominen}\ \emph {et~al.}(2017)\citenamefont
  {Suominen}, \citenamefont {Danon}, \citenamefont {Kjaergaard}, \citenamefont
  {Flensberg}, \citenamefont {Shabani}, \citenamefont {Palmstr\o{}m},
  \citenamefont {Nichele},\ and\ \citenamefont {Marcus}}]{Suominen2017}%
  \BibitemOpen
  \bibfield  {author} {\bibinfo {author} {\bibfnamefont {H.~J.}\ \bibnamefont
  {Suominen}}, \bibinfo {author} {\bibfnamefont {J.}~\bibnamefont {Danon}},
  \bibinfo {author} {\bibfnamefont {M.}~\bibnamefont {Kjaergaard}}, \bibinfo
  {author} {\bibfnamefont {K.}~\bibnamefont {Flensberg}}, \bibinfo {author}
  {\bibfnamefont {J.}~\bibnamefont {Shabani}}, \bibinfo {author} {\bibfnamefont
  {C.~J.}\ \bibnamefont {Palmstr\o{}m}}, \bibinfo {author} {\bibfnamefont
  {F.}~\bibnamefont {Nichele}},\ and\ \bibinfo {author} {\bibfnamefont {C.~M.}\
  \bibnamefont {Marcus}},\ }\bibfield  {title} {\bibinfo {title} {Anomalous
  {F}raunhofer interference in epitaxial superconductor-semiconductor
  {J}osephson junctions},\ }\href {https://doi.org/10.1103/PhysRevB.95.035307}
  {\bibfield  {journal} {\bibinfo  {journal} {Phys. Rev. B}\ }\textbf {\bibinfo
  {volume} {95}},\ \bibinfo {pages} {035307} (\bibinfo {year}
  {2017})}\BibitemShut {NoStop}%
\bibitem [{\citenamefont {Tolpygo}\ and\ \citenamefont
  {Gurvitch}(1996)}]{tolpygo_1996_critical}%
  \BibitemOpen
  \bibfield  {author} {\bibinfo {author} {\bibfnamefont {S.~K.}\ \bibnamefont
  {Tolpygo}}\ and\ \bibinfo {author} {\bibfnamefont {M.}~\bibnamefont
  {Gurvitch}},\ }\bibfield  {title} {\bibinfo {title} {Critical currents and
  {J}osephson penetration depth in planar thin-film high-t$_c$ {J}osephson
  junctions},\ }\href@noop {} {\bibfield  {journal} {\bibinfo  {journal} {Appl.
  Phys. Lett.}\ }\textbf {\bibinfo {volume} {69}},\ \bibinfo {pages} {3914}
  (\bibinfo {year} {1996})}\BibitemShut {NoStop}%
\bibitem [{\citenamefont {Beenakker}(1991)}]{PhysRevLett.67.3836}%
  \BibitemOpen
  \bibfield  {author} {\bibinfo {author} {\bibfnamefont {C.~W.~J.}\
  \bibnamefont {Beenakker}},\ }\bibfield  {title} {\bibinfo {title} {Universal
  limit of critical-current fluctuations in mesoscopic {J}osephson junctions},\
  }\href {https://doi.org/10.1103/PhysRevLett.67.3836} {\bibfield  {journal}
  {\bibinfo  {journal} {Phys. Rev. Lett.}\ }\textbf {\bibinfo {volume} {67}},\
  \bibinfo {pages} {3836} (\bibinfo {year} {1991})}\BibitemShut {NoStop}%
\bibitem [{\citenamefont {Sticlet}\ \emph {et~al.}(2020)\citenamefont
  {Sticlet}, \citenamefont {W\'ojcik},\ and\ \citenamefont
  {Nowak}}]{PhysRevB.102.165407}%
  \BibitemOpen
  \bibfield  {author} {\bibinfo {author} {\bibfnamefont {D.}~\bibnamefont
  {Sticlet}}, \bibinfo {author} {\bibfnamefont {P.}~\bibnamefont {W\'ojcik}},\
  and\ \bibinfo {author} {\bibfnamefont {M.~P.}\ \bibnamefont {Nowak}},\
  }\bibfield  {title} {\bibinfo {title} {Squid pattern disruption in transition
  metal dichalcogenide {J}osephson junctions due to nonparabolic dispersion of
  the edge states},\ }\href {https://doi.org/10.1103/PhysRevB.102.165407}
  {\bibfield  {journal} {\bibinfo  {journal} {Phys. Rev. B}\ }\textbf {\bibinfo
  {volume} {102}},\ \bibinfo {pages} {165407} (\bibinfo {year}
  {2020})}\BibitemShut {NoStop}%
\bibitem [{\citenamefont {Banerjee}\ \emph {et~al.}(2022)\citenamefont
  {Banerjee}, \citenamefont {Lesser}, \citenamefont {Rahman}, \citenamefont
  {Wang}, \citenamefont {Li}, \citenamefont {Kringhøj}, \citenamefont
  {Whiticar}, \citenamefont {Drachmann}, \citenamefont {Thomas}, \citenamefont
  {Wang}, \citenamefont {Manfra}, \citenamefont {Berg}, \citenamefont {Oreg},
  \citenamefont {Stern},\ and\ \citenamefont {Marcus}}]{Banerjee2022}%
  \BibitemOpen
  \bibfield  {author} {\bibinfo {author} {\bibfnamefont {A.}~\bibnamefont
  {Banerjee}}, \bibinfo {author} {\bibfnamefont {O.}~\bibnamefont {Lesser}},
  \bibinfo {author} {\bibfnamefont {M.~A.}\ \bibnamefont {Rahman}}, \bibinfo
  {author} {\bibfnamefont {H.~R.}\ \bibnamefont {Wang}}, \bibinfo {author}
  {\bibfnamefont {M.~R.}\ \bibnamefont {Li}}, \bibinfo {author} {\bibfnamefont
  {A.}~\bibnamefont {Kringhøj}}, \bibinfo {author} {\bibfnamefont {A.~M.}\
  \bibnamefont {Whiticar}}, \bibinfo {author} {\bibfnamefont {A.~C.~C.}\
  \bibnamefont {Drachmann}}, \bibinfo {author} {\bibfnamefont {C.}~\bibnamefont
  {Thomas}}, \bibinfo {author} {\bibfnamefont {T.}~\bibnamefont {Wang}},
  \bibinfo {author} {\bibfnamefont {M.~J.}\ \bibnamefont {Manfra}}, \bibinfo
  {author} {\bibfnamefont {E.}~\bibnamefont {Berg}}, \bibinfo {author}
  {\bibfnamefont {Y.}~\bibnamefont {Oreg}}, \bibinfo {author} {\bibfnamefont
  {A.}~\bibnamefont {Stern}},\ and\ \bibinfo {author} {\bibfnamefont {C.~M.}\
  \bibnamefont {Marcus}},\ }\bibfield  {title} {\bibinfo {title} {Signatures of
  a topological phase transition in a planar {J}osephson junction},\ }\href
  {https://arxiv.org/abs/2201.03453} {\bibfield  {journal} {\bibinfo  {journal}
  {arXiv.2201.03453}\ } (\bibinfo {year} {2022})}\BibitemShut {NoStop}%
\bibitem [{\citenamefont {Sticlet}\ \emph {et~al.}(2017)\citenamefont
  {Sticlet}, \citenamefont {Nijholt},\ and\ \citenamefont
  {Akhmerov}}]{short_junctions}%
  \BibitemOpen
  \bibfield  {author} {\bibinfo {author} {\bibfnamefont {D.}~\bibnamefont
  {Sticlet}}, \bibinfo {author} {\bibfnamefont {B.}~\bibnamefont {Nijholt}},\
  and\ \bibinfo {author} {\bibfnamefont {A.}~\bibnamefont {Akhmerov}},\
  }\bibfield  {title} {\bibinfo {title} {Robustness of {M}ajorana bound states
  in the short-junction limit},\ }\href
  {https://doi.org/10.1103/PhysRevB.95.115421} {\bibfield  {journal} {\bibinfo
  {journal} {Phys. Rev. B}\ }\textbf {\bibinfo {volume} {95}},\ \bibinfo
  {pages} {115421} (\bibinfo {year} {2017})}\BibitemShut {NoStop}%
\bibitem [{\citenamefont {Peierls}(1933)}]{Peierls1933}%
  \BibitemOpen
  \bibfield  {author} {\bibinfo {author} {\bibfnamefont {R.}~\bibnamefont
  {Peierls}},\ }\bibfield  {title} {\bibinfo {title} {Zur theorie des
  diamagnetismus von leitungselektronen},\ }\href
  {https://doi.org/10.1007/BF01342591} {\bibfield  {journal} {\bibinfo
  {journal} {Zeitschrift f{\"u}r Physik}\ }\textbf {\bibinfo {volume} {80}},\
  \bibinfo {pages} {763} (\bibinfo {year} {1933})}\BibitemShut {NoStop}%
\bibitem [{\citenamefont {Hofstadter}(1976)}]{PhysRevB.14.2239}%
  \BibitemOpen
  \bibfield  {author} {\bibinfo {author} {\bibfnamefont {D.~R.}\ \bibnamefont
  {Hofstadter}},\ }\bibfield  {title} {\bibinfo {title} {Energy levels and wave
  functions of bloch electrons in rational and irrational magnetic fields},\
  }\href {https://doi.org/10.1103/PhysRevB.14.2239} {\bibfield  {journal}
  {\bibinfo  {journal} {Phys. Rev. B}\ }\textbf {\bibinfo {volume} {14}},\
  \bibinfo {pages} {2239} (\bibinfo {year} {1976})}\BibitemShut {NoStop}%
\bibitem [{\citenamefont {Ando}(1991)}]{disorder}%
  \BibitemOpen
  \bibfield  {author} {\bibinfo {author} {\bibfnamefont {T.}~\bibnamefont
  {Ando}},\ }\bibfield  {title} {\bibinfo {title} {Quantum point contacts in
  magnetic fields},\ }\href {https://doi.org/10.1103/PhysRevB.44.8017}
  {\bibfield  {journal} {\bibinfo  {journal} {Phys. Rev. B}\ }\textbf {\bibinfo
  {volume} {44}},\ \bibinfo {pages} {8017} (\bibinfo {year}
  {1991})}\BibitemShut {NoStop}%
\bibitem [{\citenamefont {Groth}\ \emph {et~al.}(2014)\citenamefont {Groth},
  \citenamefont {Wimmer}, \citenamefont {Akhmerov},\ and\ \citenamefont
  {Waintal}}]{kwant}%
  \BibitemOpen
  \bibfield  {author} {\bibinfo {author} {\bibfnamefont {C.~W.}\ \bibnamefont
  {Groth}}, \bibinfo {author} {\bibfnamefont {M.}~\bibnamefont {Wimmer}},
  \bibinfo {author} {\bibfnamefont {A.~R.}\ \bibnamefont {Akhmerov}},\ and\
  \bibinfo {author} {\bibfnamefont {X.}~\bibnamefont {Waintal}},\ }\bibfield
  {title} {\bibinfo {title} {Kwant: a software package for quantum transport},\
  }\href {https://doi.org/10.1088/1367-2630/16/6/063065} {\bibfield  {journal}
  {\bibinfo  {journal} {New J. Phys.}\ }\textbf {\bibinfo {volume} {16}},\
  \bibinfo {pages} {063065} (\bibinfo {year} {2014})}\BibitemShut {NoStop}%
\end{thebibliography}
\end{document}